\documentclass[utf8,usenatbib]{frontiersSCNS} 
\pdfoutput=1
\usepackage{url,hyperref,lineno,microtype,subcaption}
\usepackage[onehalfspacing]{setspace}
\usepackage{datetime}
\usepackage{fmtcount}
\usepackage{etoolbox}
\usepackage{fcprefix}
\usepackage{booktabs}
\usepackage[normalem]{ulem}
\usepackage{soul}
\usepackage{amssymb}
\usepackage{xcolor}
\usepackage{placeins}

\definecolor{granata}{HTML}{B30033}

\def\keyFont{\fontsize{8}{11}\helveticabold }
\def\firstAuthorLast{Millas {et~al.}} 
\def\Authors{Dimitrios Millas\,$^{1,4,*}$, Maria Elena Innocenti\,$^{1,*}$, Brecht Laperre\,$^{1}$, Joachim Raeder\,$^{2}$, Stefaan Poedts\,$^{1,3}$ and Giovanni Lapenta\,$^{1}$}
\def\Address{$^{1}$Centre for mathematical Plasma Astrophysics, Department of Mathematics, KU Leuven, Leuven, Belgium

$^{2}$ Institute for the Study of Earth, Oceans and Space, University of New Hampshire, Durham, NH, USA

$^{3}$ Institute of Physics, University of Maria Curie-Sk{\l}odowska, Lublin, Poland

$^{4}$ Department of Physics and Astronomy, University College London, London, UK}

\def\corrAuthor{Dimitrios Millas}
\def\corrEmail{dimitrios.millas@kuleuven.be}

\begin{document}
\onecolumn
\firstpage{1}

\title[DOI analysis for space weather]{Domain of Influence analysis: implications for Data Assimilation in space weather forecasting} 

\author[\firstAuthorLast ]{\Authors} 
\address{} 
\correspondance{} 

\extraAuth{Maria Elena Innocenti \\ mariaelena.innocenti@kuleuven.be} 

\maketitle

\begin{abstract}
Solar activity, ranging from the background solar wind to energetic coronal mass ejections (CMEs), is the main driver of the conditions in the interplanetary space and in the terrestrial space environment, known as space weather. A better understanding of the Sun-Earth connection carries enormous potential to mitigate negative space weather effects with economic and social benefits. Effective space weather forecasting relies on data and models. In this paper, we discuss some of the most used space weather models, and propose suitable locations for data gathering with space weather purposes. 
We report on the application of \textit{Representer analysis (RA)} and \textit{Domain of Influence (DOI) analysis} to three models simulating different stages of the Sun-Earth connection: the OpenGGCM and Tsyganenko models, focusing on solar wind - magnetosphere interaction, and the PLUTO model, used to simulate CME propagation in interplanetary space. Our analysis is promising for space weather purposes for several reasons. First, we obtain quantitative information about the most useful locations of observation points, such as solar wind monitors. For example, we find that the absolute values of the DOI are extremely low in the magnetospheric plasma sheet. Since knowledge of that particular sub-system is crucial for space weather, enhanced monitoring of the region would be most beneficial. Second, we are able to better characterize the models. Although the current analysis focuses on spatial rather than temporal correlations, we find that time-independent models are less useful for Data Assimilation activities than time-dependent models. Third, we take the first steps towards the ambitious goal of identifying the most relevant heliospheric parameters for modelling CME propagation in the heliosphere, their arrival time, and their geoeffectiveness at Earth.

\tiny
 \keyFont{ \section{Keywords:} solar wind, coronal mass ejections (CMEs), magnetohydrodynamics (MHD), numerical simulations, statistical tools, domain of influence, observations} 
\end{abstract}

\section{Introduction}

Solar activity affects the terrestrial environment with a constantly present but highly variable solar wind and with higher energy, transient events, such as flares and Coronal Mass Ejections (CMEs).

``Space weather''~\citep{bothmer2007space} is the discipline that focuses on the impact of these solar drives on the solar system and in particular on the Earth and its near space environment.

Space weather events can have serious effects on the health of astronauts and on technology, with potentially large economic costs~\citep{Eastwood2017}.  The importance of space weather forecasting has grown with societal dependence on advanced space technology, on communication and on the electrical grid. For example, the Halloween 2003 solar storms that impacted Earth between 19th of October 2003 and 7th of November 2003 caused an hour long power outage in Sweden~\citep{pulkkinen2005geomagnetic}, forced airline flight reroutes, and affected communication and satellite systems \citep{plunkett2005extreme}. The ``great geomagnetic storm" of March 13-14, 1989 caused, among other disruptions, a blackout of up to nine hours in most of Quebec Province, due to a massive failure experienced by the power grid Hydro-Quebec Power Company~\citep{allen1989effects}.

In order to improve space weather forecasting, accurate models of the Sun-Earth connection are needed. Such forecasts are challenging because of the complexity of the processes involved and the large range of spatial and temporal scales.  Commonly the heliosphere is divided into sub-systems, where each one is simulated with a different model, such that the models feed into each other~\citep{luhmann2004coupled, toth2005space}. These models can be either physics-based or empirical. Empirical models (such as, in the solar domain,~\citet{altschuler1969magnetic, schatten1969model, schatten1971current, wang1992potential,nikolic2019}) usually require less computational resources, enabling faster forecasts. They can also serve as a baseline for physics-based models~\citep{siscoe2004roles}. However, empirical models lack the sophistication of more expensive first-principles based numerical models. Recently, machine learning methods have emerged, that can provide a new approach to space weather forecasting~\citep{camporeale2019challenge, laperre2020dynamic}. Most of these methods, while promising, must still undergo extensive validation.

The technique of Data Assimilation (DA) was developed to improve model predictions by properly initializing models from data and by keeping a model on track during its time evolution. \citep{kalnay2003atmospheric,bouttier1999data,evensen2009data}.
DA methods were originally applied to atmosphere and ocean models, which exhibit a large degree of inertia.  The latter is also true for the solar wind, but not for the magnetosphere-ionosphere system, which is strongly driven and dissipative.
Therefore, in space weather forecasting, DA aims not only at initializing the models, but also  at using information from various observations to bring the evolution of a system as predicted from a model closer to the real system evolution~\citep{kalman1960new,bouttier1999data,bishop2001introduction,evensen2009data,le1986variational, innocenti2011improved}, making up for model deficiencies in the terms of resolution and incomplete physical description. 

The quantity and quality of available data is a critical factor in the effectiveness of Data Assimilation. This is the reason why fields where data can be obtained more easily and continuously have shown early successes in DA implementations. Examples of these fields are meteorology and oceanography, and, in space sciences, ionospheric and radiation belt physics~\citep{bennett1992inverse,ghil1991data,egbert1996data,kalnay2003atmospheric,rigler2004adaptive,schunk2004global,kondrashov2007kalman}.
Examples of DA applications targeting specifically the interplanetary space environments are~\citet{schrijver2003photospheric, mendoza2006data, arge2010air, innocenti2011improved, skandrani2014flip, lang2019variational, Lang2019,Lang2020}.

Representer analysis (RA) and Domain of Influence analysis (DOI) \citep{bennett1992inverse, egbert1996data,echevin2000horizontal, evensen2009data,skandrani2014flip}, briefly summarized in Section~\ref{sec: Methods}, are powerful statistical tools used to estimate the effectiveness of DA techniques when applied to a specific model, without assimilating actual data.
Such analysis can be used in several ways. It allows us
to optimize assimilation strategies, it may uncover model biases 
that can then be addressed by further model development,  and it may be used to optimize the observation systems that provide operational data for DA.   For example, RA/DOI can be used to optimize locations for solar wind monitors, such as locations proposed near the L5 Lagrangian point \citep{Vourlidas2015,Lavraud2016,Pevtsov2020}. 

In the present paper, the RA and DOI analysis is applied to three models: the OpenGGCM magnetosphere - ionosphere model (Section~\ref{sect:OpenGGCM}), two of the empirical Tsyganenko magnetosphere magnetic field models (Section~\ref{sect:Tsyg}), and a solar wind simulation based on the PLUTO code (Section~\ref{sect:PLUTO}). These models simulate critical sub-systems in the Sun-Earth connection with a focus on the terrestrial magnetosphere and Coronal Mass Ejection propagation. 

The present paper provides insights into the locations of the terrestrial magnetosphere that should be prioritized (ideally, in absence of orbital constraints) for space weather forecasting and monitoring activity. We compare a time-dependent, physics based model (e.g., OpenGGCM) and time-independent, empirical (e.g., Tsyganenko) models in terms of the expected benefits that DA can provide. We conclude that time-dependent models should be preferentially chosen for DA. We take the first steps towards the goal if understanding the main physical parameters, close to the Sun and in interplanetary space, that control CME propagation and hence their arrival time at Earth.

This manuscript is organized as follows: in Section \ref{sec: Methods} we introduce the theoretical background on RA and the DOI; Section \ref{sec: Applications} discusses the application of the method to the different models; in Section \ref{sec: Conclusions} we summarize the results and discuss potential improvements and new applications.

\section{Representer analysis and Domain of Influence analysis}\label{sec: Methods}

This Section introduces the mathematical basis of RA and DOI analysis. The reader is referred to~\citet{skandrani2014flip} and references therein for an in depth derivation.

Let us start from a system described by the state variable vector $\mathbf{x}_t \in\;  \mathbb{R}^n$.$\mathbf{x}_t$ is a vector containing the $n$ state variables that describe the system at a time $t$. Let us assume that the evolution of the system can be described as a discrete-time process controlled by an evolution law $\mathbf{A}$. The state of the system then evolves as follows: $\mathbf{x}_t=\mathbf{A} (\mathbf{x}_{t-1}) + \mathbf{w}_{t-1}$, where  $\mathbf{w} \in\;  \mathbb{R}^n$ is process noise. The process noise is assumed to be Gaussian and with covariance matrix Q. 

If a model $\mathbf{M}$ (for example, a simulation model) of the evolution law $\mathbf{A}$ is available, we can obtain, following~\citet{kalman1960new, evensen2009data}, a prior estimate $\hat{\mathbf{x}}^-_t$ of the state variable $\mathbf{x}_t$ through the simulation model as
\begin{equation}
\centering
    \hat{\mathbf{x}}_t^- = \mathbf{M} (\hat{\mathbf{x}}^-_{t-1}) + \mathbf{w}_{t-1}.
  \label{eq:mod}
\end{equation}

Assume now that we have $m$ observational values or measurements $\mathbf{z}_t \in \mathbb{R}^m$. These measurements can be mapped to the current state $\mathbf{x_t}$ through the so-called observation operator $\mathbf{H} \in \mathbb{R}^{m \times n}$, such that $\mathbf{z}_t=\mathbf{H} \mathbf{x}_t+\mathbf{\nu}_t$.
Here, $\mathbf{\nu}$ is the (assumed Gaussian) measurement noise, with a covariance matrix $\mathbf{R}$.

It can be then shown~\citep{bishop2001introduction} that a posterior estimate of the state ($\hat{\mathbf{x}}_t$) can be obtained from the prior estimate of the state ($\hat{\mathbf{x}}_t^{-}$), obtained from Eq.~\ref{eq:mod}, as follows:
\begin{equation}
\centering
\hat{\mathbf{x}}_t = \hat{\mathbf{x}}_t^{-} + \mathbf{K}_t \left(\mathbf{z}_t - \mathbf{H}  \hat{\mathbf{x}}_t^{-} \right).
\label{eq:corr1}
\end{equation}
Here, the term $\left(\mathbf{z}_t - \mathbf{H}  \hat{\mathbf{x}}_t^{-} \right)$ is called the ``innovation”, and represents the difference between the measurements and their expected values, calculated by applying the observation operator to the prior state estimate. The Kalman gain $\mathbf{K}_t$ is obtained by minimizing the posterior error covariance matrix. This is the ``correction” step of the Kalman filter, where the Kalman gain is calculated and the  estimate and error covariance matrix of the posterior state are updated. The ``prediction” (forecast) phase of the filter results in the calculation of the prior state estimate and prior error covariance matrix (used to compute the Kalman gain).

The prior and posterior error covariance matrices are respectively defined as
\begin{equation}
    \mathbf{P}_t^-= \mathbb{E}\left[ \left(\mathbf{x}_t - \hat{\mathbf{x}}_t^{-} \right) \left(\mathbf{x}_t - \hat{\mathbf{x}}_t^{-} \right)^T \right], \quad \mathbf{P}_t= \mathbb{E}\left[ \left(\mathbf{x}_t - \hat{\mathbf{x}}_t \right) \left(\mathbf{x}_t - \hat{\mathbf{x}}_t \right)^T \right],
\end{equation}
where $\mathbb{E}$ is the expected value, $\mathbf{x}_t$ is the ``real”, unknown system state and $\epsilon^-= \left( \hat{\mathbf{x}}_t^{-}- {\mathbf{x}}_t  \right)$ and $\epsilon= \left( \hat{\mathbf{x}}_t- {\mathbf{x}}_t  \right)$ are the prior and posterior errors, calculated as the difference between the prior ($\hat{\mathbf{x}}_t^{-}$)/posterior ($\hat{\mathbf{x}}_t$) state and the real state, $\mathbf{x}_t$. Notice that, although these are the definitions of the error covariances, this is not how they are computed in the filter, since the real state is not known.

The formula for the calculation of the posterior state, Eq.~\ref{eq:corr1}, can be written as
\begin{equation}
\centering
\hat{\mathbf{x}}_t = \hat{\mathbf{x}}_t^{-} + \mathbf{r} \mathbf{b}
\label{eq:corr2}
\end{equation}
where $\mathbf{r} \in\;  \mathbb{R}^{n \times m}$ and $\mathbf{b} \in\;  \mathbb{R}^{m}$ are defined as  

\begin{equation}
\centering
\mathbf{r}= \mathbf{P}^{-} \mathbf{H}^T, \qquad \mathbf{b}=\left( \mathbf{H} \mathbf{P}^{-} \mathbf{H}^{T}+\mathbf{R} \right)^{-1} \left(\mathbf{z}- \mathbf{H} \hat{\mathbf{x}}^{-} \right),
\label{eq:corr2_2}
\end{equation}
with $\mathbf{R}$ the measurement noise covariance matrix. The time index $t$ has been dropped for ease of reading. 

We will from now on assume that the system (and in particular, the observation operator $\mathbf{H}$) is linear. Then, each column of the matrix $\mathbf{r}$, denoted as $\mathbf{r}_j$ with $j=1, \dots, m$, is the representer associated to a given observation~$z_j$ (remember that $\mathbf{z}$ is the vector with $m$ observations), and gives a measure of the impact of that observation in “correcting” the prior state estimate. If we further assume that each observation $j$ is located at grid point $k_j$, and is associated to a state variable, then each column $\mathbf{r}_j$ (now $\mathbf{r}_{k_j}$, given the assumption mentioned above) contains the covariances (“cov”) between the prior errors at the observation point $k_j$ and at every other grid node, for all the state variables.

Since the real state is not available for error covariance calculations, an ensemble of simulations can be used to estimate the prior errors instead. An ensemble~\citep{evensen2009data} can be generated by perturbing one or several of the sources of model errors. In this work, ensembles are generated for each model by perturbing the respective initial / boundary conditions. 
Once the ensemble is available, the covariances of the prior errors at a certain simulated time can be approximated as the ensemble covariances (``$\text{cov}^{ens}$") of the prior errors. These in turn become the ensemble covariances of the simulated state variables, if one assumes that the prior errors are unbiased. The ensemble covariance between the state variable x and y is defined as
\begin{equation}
\text{cov}^{ens}(x, y) = \frac{1}{N} \sum^N_{s=1} \left[ \left( x^-_s - \frac{1}{N} \sum^N x_s^- \right) \left( y_s^- - \frac{1}{N} \sum^N_{s=1} y_s^- \right) \right]
\end{equation}
where $N$ is the number of members in the ensemble, and $x$ and $y$ represent two state variables (notice that, for ease of reading, we indicate here two state variables as x and y, while earlier we indicated as vector $\mathbf{x}$ all the state variables). 

In a set of simulations, the prior state variables are the simulation results at a specific time. Being able to calculate the representers associated to observations from the ensemble rather than from assimilating observations simplifies the RA significantly. 

Following the discussion of the representer term $\mathbf{r}$ in the calculation of the posterior state, Eq.~\ref{eq:corr2} and Eq.~\ref{eq:corr2_2}, we now examine the term $\mathbf{b}$.
Assuming that the only observation point for observation $j$ is at grid node $k_j$ (i.e., the row $j$ of the matrix $\mathbf{H}$ has only the term $k_j$ different from zero), the element of $\mathbf{b}$ associated to the observation at grid point $k_j$, denoted as $b_{k_j}$, becomes, from Eq.~\ref{eq:corr2_2}:
\begin{equation}
b_{k_j}= \dfrac{z_{k_j}- \hat{x}^-_{k_j}}{cov\left(\epsilon^-_{k_j},\epsilon^-_{k_j}\right) + cov\left(\epsilon^z_{k_j},\epsilon^z_{k_j}\right)},
\label{eq:bkj}
\end{equation}
where we have made use of the simplified form of the matrix $\mathbf{H}$ and where $\epsilon^z_{k_j}$ is the observation error associated with the observation $z_{k_j}$.

If $x_i$ is one of the state variables at grid node $i$, the correction to $x_i$ brought by the assimilation of the measure $z_{k_j}$, following Eq.~\ref{eq:corr2}, Eq.~\ref{eq:corr2_2}, Eq.~\ref{eq:bkj} and some straightforward manipulation based on the definitions of covariance, variance, correlation, can be written as \citep{skandrani2014flip}

\begin{equation}
\centering
\hat{x_i} - \hat{x}_i^{-}= \text{corr}^{ens}\left(\hat{x}_{k_j}^-, \hat{x}_i^{-} \right) F\left(z_{k_j} \right). 
\label{eq:correction}
\end{equation}

Here, $F(z_{k_j})$ is the modulation factor and $\text{corr}^{ens}\left(\hat{x}_{k_j}^-, \hat{x}_i^{-} \right)$ the correlation. The correlation is computed from the ensemble, and is calculated between the state variable at node $k_j$ and at node $i$. This correlation reflects how a change at node $k_j$, caused e.g. by the assimilation of the measurement $z_{k_j}$, will influence the node $i$, and is what we call the DOI. The correlation over the ensemble is defined, using the dummy variables $x$ and $y$ for brevity, as:
\begin{equation}
\centering
\text{corr}^{ens}(x,y) = \frac{\text{cov}^{ens}\left(x, y \right) } {\sqrt{\text{var}\left( x \right)\text{var}\left( y\right)}}.
\end{equation}

The modulation factor $F(z_{k_j})$ in Eq.~\ref{eq:correction} depends, among other things, on the measurement $z_{k_j}$ and on the error associated to the measure $z_{k_j}$. Hence Data Assimilation has to be  performed to calculate this term. The $\text{corr}^{ens}\left(\hat{x}_{k_j}^-, \hat{x}_i^{-} \right)$ term reflects how large we can expect the area that will be affected by the assimilation of $z_{k_j}$ to be. But to know how large the difference between the posterior and prior state, $\hat{x_i} - \hat{x}_i^{-}$, will be, we need to know the modulation factor as well.

So now the DOI of the measurement $z_{k_j}$ on the state variable at grid point $i$, $x_i$, can be defined as  
\begin{equation}
\centering
DOI(z_{k_j}, x_{i})= \text{corr}^{ens}\left(\hat{x}_{k_j}^-, \hat{x}_i^{-} \right).
\end{equation}
One can see from its definition that the DOI can be calculated \textit{before assimilation} by computing the ensemble correlation of the state variable value at the grid point $k_j$ with that at grid point $i$.
Dropping the $i$ index, i.e. examining the expected impact of measurement $z_{k_j}$ on all the state variables $\mathbf{x}$ at all grid points, we obtain the more general definition of the DOI as
\begin{equation}
\centering
DOI(z_{k_j})= \text{corr}^{ens}\left(\hat{x}_{k_j}^-, \hat{\mathbf{x}}^{-} \right).
\end{equation}

We can then draw ``DOI maps" that show the correlation between a field at grid point $k_j$, the ``observation point", and the other grid points.

Notice that in this derivation we have assumed, for simplicity, that the measurement $z$ and the state variable $x$ refer to the same field, for example, the x component of the magnetic field, or of the velocity. This simplifies the formulation of the numerator of Eq.~\ref{eq:bkj} and improves the readability of the derivation. \citet{skandrani2014flip} shows examples where the DOI is calculated between different fields, e.g. magnetic field and velocity.

DOI analysis has the advantage that it can be calculated for all state variables and at any grid point without actual assimilation, i.e. without the need for measurements $\mathbf{z}$. In order to compute the DOI at a time step $t$, we only require evolving the ensemble up to said time step $t$, and then performing the correlation over the ensemble between the state variable value at the observation point $k_j$ and at all other grid points.

Because DOI values are derived from a correlation they are bounded between -1 and 1. $|DOI| \sim 1$ indicates that the field at that specific point significantly changes when the same field (or a different field, in the case of cross-correlation) varies at the observation point. $|DOI| \sim 0$ indicates the opposite, i.e., variation at the observation point have little or no effect. Thus, DOI analysis also provides information on how information propagates within the model, and therefore sheds light on the physical processes within the model.  We will exploit this property in Section~\ref{sect:Tsyg}.

We note that in this study we only focus on spatial correlations, neglecting temporal correlations. In other words, the following analysis (i.e. the calculation of variances, DOI, etc.) is restricted to specific instances in time, rather than examining correlations between fields at difference times as well. The dependence on time will be addressed in a future project.

The RA is applied to ``artificial'' data, obtained from ensembles of simulations focused on different processes of interest in the Sun-Earth connection: the interaction of the solar wind with the terrestrial magnetosphere (via OpenGGCM and Tsyganenko simulations) and the propagation of a CME-like event in the steady solar wind (PLUTO). 

OpenGGCM and the Tsyganenko Geomagnetic Field Models both simulate the interaction of the solar wind with the magnetospheric system. OpenGGCM is a physics-based magnetohydrodynamic (MHD) model, while the Tsyganenko models are semi-empirical best-fit representations for the magnetic field, based on a large number of satellite observations \citep{tsyganenko1995modeling, tsyganenko2002modela, tsyganenko2005modeling}.

PLUTO is an MHD-modelling software used to simulate the propagation of a CME in the background solar wind. This software can be used to numerically solve the partial differential equations encountered in plasma physics problems, in conservative form, in different regimes (from hydrodynamics to relativistic MHD). The structure of the software is explained in \cite{Mignone2007,Mignone2012}. Full documentation and references can be found in the relevant web page: \url{http://plutocode.ph.unito.it/}.

Because the DOI analysis is an ensemble based technique the size of the ensemble and its properties matter. In order to test for sufficient size, we performed the DOI calculation using a limited, random subset of the ensemble, which we gradually expanded. We found that using at 25 runs were sufficient to obtain a consistent ensemble mean, variance, and DOI. We note, however, that this may change for different choices of simulation resolution and parameters used for the generation of the ensemble.

\section{Applications}\label{sec: Applications}
\subsection{Magnetospheric applications I: OpenGGCM} \label{sect:OpenGGCM}

The OpenGGCM (Open Geospace General Circulation Model) is a MHD based model that simulates the interaction of the solar wind with the magnetosphere-ionosphere-thermosphere system. OpenGGCM is available at the Community Coordinated Modeling Center at NASA/GSFC for model runs on demand (see: http://ccmc.gsfc.nasa.gov). This model has been developed and continually improved over more than two decades. Besides numerically solving the MHD equations with high spatial resolution in a large volume containing the magnetosphere, the model also includes ionospheric processes and their electrodynamic coupling with the magnetosphere. 

The mathematical formulation of the software is described in \citet{raeder03t}.
The latest version of OpenGGCM, used here, is coupled with the Rice Convection Model (RCM), \citep{toffoletto2001modeling}, which treats the inner magnetosphere drift physics better than MHD and allows for more realistic simulations that involve the ring current \citep{cramer2017plasma}. The model is both modular and efficiently parallelized using the message passing interface (MPI).  It is written in Fortran and C, and uses extensive Perl scripting for pre-processing.  The software runs on virtually any massively parallel supercomputer available today.  

OpenGGCM uses a stretched Cartesian grid \citep{raeder03t} that is quite flexible.  There is a minimal useful resolution, about 150x100x100 cells, that yields the main magnetosphere features but does not resolve mesoscale structures such as FTEs or small plasmoids in the tail plasma sheet.  At the other end, we have run simulations with grids as large as $\sim$1000$^3$ (on some 20,000 cores).  In terms of computational cost that is almost a 10$^4$ ratio.  Here, we used a grid of 325x150x150 cells which is sufficient for the purposes of this study and runs faster than real time on a modest number of cores.

 OpenGGCM has been used for numerous studies of magnetospheric phenomena such as storms \citep{raeder01a,Raeder2005PO,Connor2016MO}, substorms \citep{raeder00c,Ge2011b,raeder2010a}, magnetic reconnection \citep{Dorelli2004AN,Raeder2006FL,berchem95l}, field-aligned currents \citep{moretto2005,vennerstroe2005,Raeder2017,Anderson2017CO}, and magnetotail processes \citep{Zhu2009IN,Zhou2012DI,Shi2014SO}, to name a few.
  
The boundary conditions require the specification of the three components of the solar wind velocity and magnetic field, the plasma pressure and the plasma number density at 1 AU, which are obtained from ACE observations~\citep{stone1998advanced} and applied for the entire duration of the simulation at the sunward boundary.

We generate an ensemble of 50 OpenGGCM simulations by perturbing the $v_x$ component of the input solar wind velocity. Changing this particular parameter guarantees a direct and easy way to interpret magnetospheric response. 

First, we run a reference simulation using the observed solar wind values at 1 AU starting from May 8$^{th}$, 2004, 09:00 UTC (denoted as $t_0$) until 13:00 UTC on May 8$^{th}$, 2004. We choose this period because it is relatively quiet: no iCME were registered in the Richardson/ Cane list of near-Earth interplanetary CMEs~\citep{richardson2010near} and, as it is common during the declining phase of the solar cycle, geomagnetic activity is driven by Corotating Interaction Regions and High Speed Streams~\citep{tsurutani2006corotating}. The study of outlier events, such as CME arrival at Earth, is left as future work.

To generate the ensemble, the solar wind compression is changed in each of the ``perturbed" simulations. The perturbed velocity in the $x$ (Earth-to-Sun) direction is set, for the entire duration of each simulation, to a \textit{constant} value obtained by multiplying the average observed $v^{avg}_x$ by a random number $S$ sampled from a normal distribution with mean $\mu=1$ and standard deviation $\sigma = 0.1$: 
\begin{equation}
    v_x = S v^{avg}_x, \text{ with } S \in \mathcal{N}(1,(0.1)^2).
\end{equation}
The time period we use to calculate the average is the duration of the reference simulation, between 9:00 UTC and 13:00 UTC on May 8$^{th}$, 2004.

Our choices for the generation of the ensemble are determined by the necessity to preserve both the Gaussian characteristic of the model error and the physical significance of the simulations: the solar wind compression in all ensemble members is is not too far from the reference value. With such low standard deviation, the average of the obtained perturbed value plus/minus several sigmas are still within the typical range for the solar wind: the minimum and maximum values of the constant, perturbed input velocities are $|v_x | \sim$ 363 km/s and $|v_x| \sim$  583 km/s respectively. The solar wind velocity $v_x$ is negative in the Geocentric Solar Ecliptic (GSE) coordinates used here. The $v_x$ values that we obtain in this way are not supposed to be representative of the full range of values that $v_x$ can assume; they are used to generate an ensemble of simulations ``slightly perturbed" with respect to our reference simulation. We note that the real distribution of solar wind velocity is far from a normal distribution, with two distinct peaks and extreme outliers, and would not be appropriate to produce the required ensemble. We refer the reader to \citet{fortin2014should} for optimal procedures on how to choose the variance of the ensemble.

The ensemble analysis requires running 50 simulations to produce the ensemble members, plus the unperturbed reference simulation. Each run  takes $\sim$ 12 hours on 52 cores on the supercomputer Marconi-Broadwell (Cineca, Italy), for a total cost of $\sim$ 32000 core hours.

\begin{figure}[h!]
\begin{center}
\includegraphics[width=0.9\textwidth]{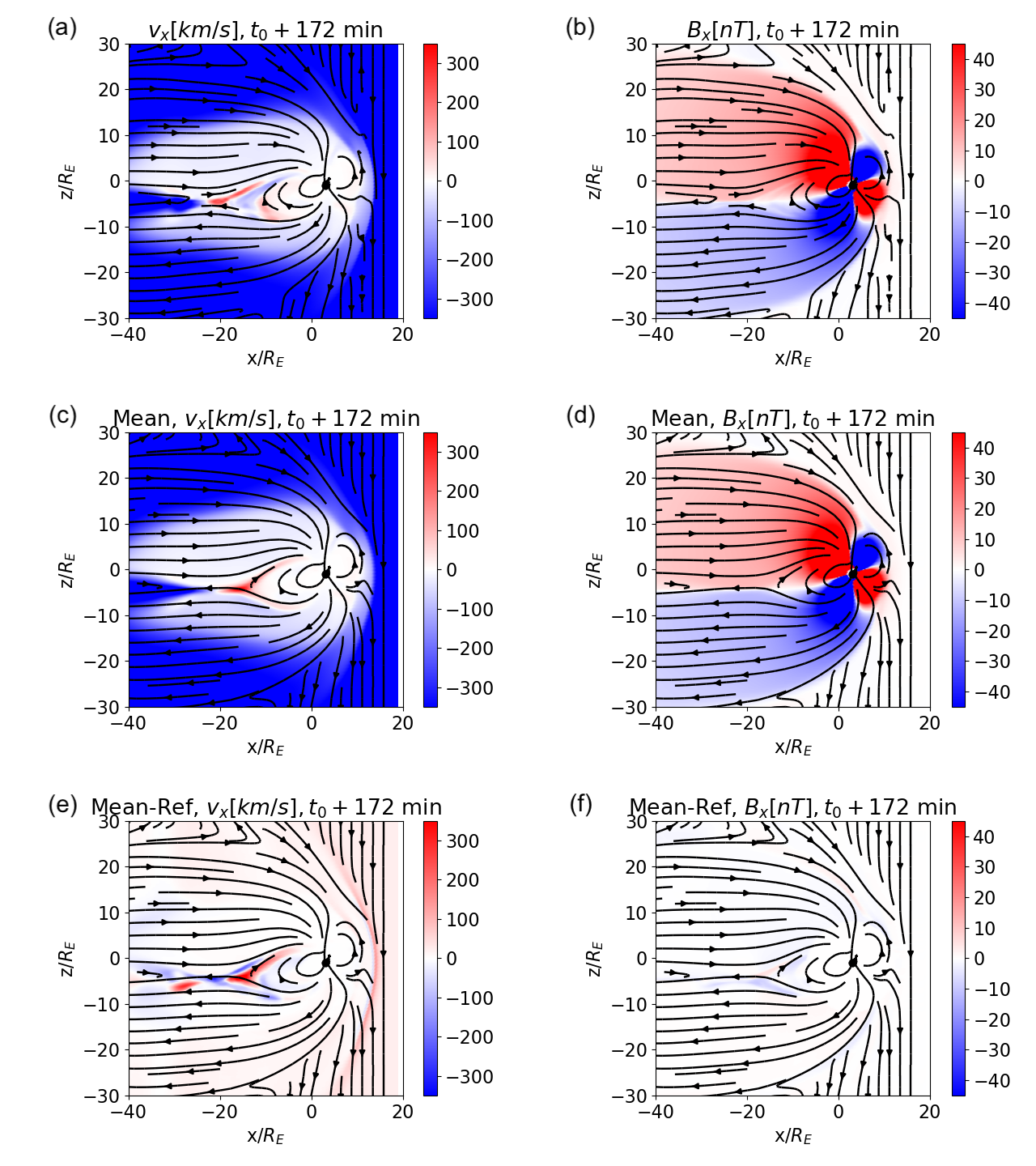}
\end{center}
\caption{Reference simulation (panel (a) and (b)), ensemble mean (panel (c) and (d)) and difference between the ensemble mean and the reference simulation results (panel (e) and (f)) for the ensemble of OpenGGCM magnetospheric simulations. The ensemble is generated by perturbing the $v_x$ solar wind boundary condition. $v_x$ is depicted in panel (a), (c), (e), $b_x$ in (b), (d), (f). The coordinate system is GSE The depicted time is 172 minutes after the beginning of the simulation, May 8$^{th}$ 2004, 9:00 UTC. The boundary conditions at the sunward boundary of the ``reference" simulation are observed solar wind values.}
\label{fig:OpenGGCM_1}
\end{figure}

We verify that the prior errors are unbiased (as assumed in the derivation of the method summarized in Section~\ref{sec: Methods}) by comparing the reference simulation and the average of the ensemble. We note that the ensemble mean is an appropriate metric to use in this case because the perturbed simulations have not been generated in order to represent all possible solar wind velocity values, but small perturbations around the reference case.

Figure~\ref{fig:OpenGGCM_1} shows this comparison in the xz plane for the x component of the velocity and of the magnetic field at a fixed time, (172 minutes after the beginning of the simulation), for both the reference simulation (panel (a) and (b)) and the average of the ensemble (panel (c) and (d)). The magnetic field lines, depicted in black, are calculated from the reference simulation in panel (a) and (b) and from the ensemble average in panel (c) and (d). The distances are normalized by the Earth radius $R_E$.

Visual inspection of panel (a), (b), (c), (d) and of the difference between the ensemble mean and the reference simulation results, depicted in panels (e) and (f) for $v_x$ and $B_x$ respectively, highlight the areas where the behavior differs most within the ensemble: the bow shock, the plasma sheet, and the neutral line.  The former is a plasma discontinuity that moves back and forth in response to the changing solar wind Mach number, and thus gets smeared out in the ensemble.  The latter is a region of marginal stability in the magnetosphere that reacts in a non-linear way to solar wind changes.

In order to determine if 50 ensemble members are sufficient for our analysis, we have compared corresponding plots of the difference between the ensemble mean and the reference simulation for $v_x$ with decreasing number (40, 30, 20) of ensemble members. 
We find that, with decreasingly smaller ensembles, the plasma sheet structure seen in Fig.~\ref{fig:OpenGGCM_1}, panel (e), is only minimally affected. However, the differences around the bow shock become more pronounced. The velocity difference increases in the solar wind and magnetosheath as well, and the magnetic field structure at the magnetosphere/ solar wind interface (as shown by the magnetic field lines, which are drawn for the average field in panel (e) and (f) and similar analysis) begin to change significantly with respect to the reference simulation. By comparing the plots with 50, 40, 30 and 20 ensemble members, we conclude that 30 is the minimum number of ensemble members that gives average fields compatible with the reference simulation, with our choice of perturbation to generate the ensemble.

Panels (a) and (b) of Figure~\ref{fig:OpenGGCM_1} show characteristic signatures of magnetic reconnection in the magnetotail, i.e, the X pattern and the formation of dipolarization fronts in the magnetic field lines, and the presence of earthward and tailward jets in $v_x$ departing from the X point. We provide a movie showing the dynamic evolution in the supplementary material (ReferenceSimVx.avi). The movie shows the solar wind $b_z$ time variation and the subsequent occurrence of several magnetopause/ magnetotail reconnection events. The ``formation" of the magnetosphere occurs during the first $\sim 30$ minutes of the simulation and should be disregarded.

The movie MeanVx.avi, also in the supplementary material, shows how the global evolution changes in the ensemble mean: the magnetopause and magnetotail reconnection patterns are still overall visible, but smoothed out by the averaging procedure with respect to the reference simulation, since the different ensemble instances reconnect at different times and the smaller scale features of each single run are averaged away.

\begin{figure}[h!]
\begin{center}
\includegraphics[width=0.9\textwidth]{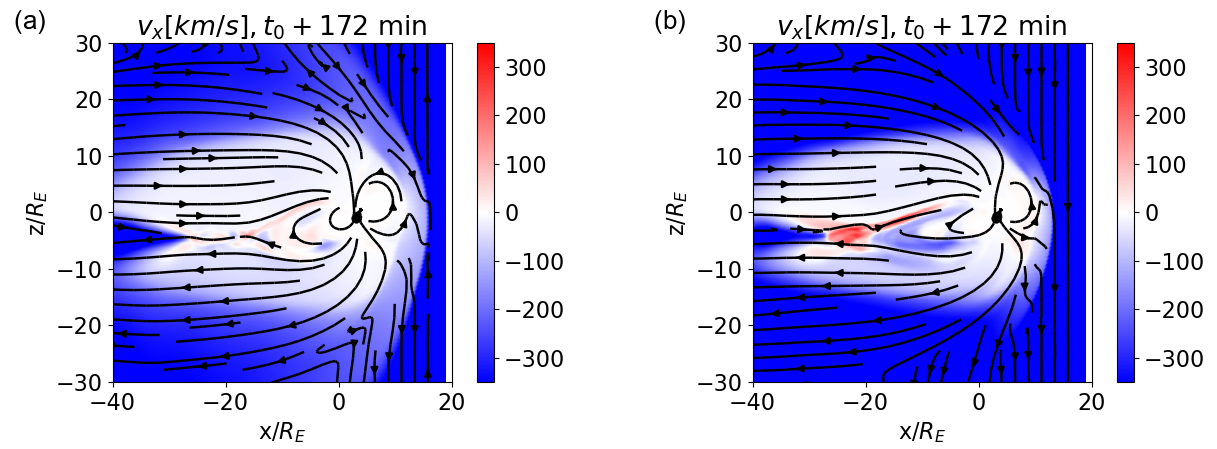}
\end{center}
\caption{$v_x$ field component, at the same time as Figure~\ref{fig:OpenGGCM_1}, for the ensemble member with the lowest (panel (a)) and highest (panel (b)) absolute value of the perturbed, inflowing $v_x$ velocity component in the OpenGGCM ensemble.}
\label{fig:HighLowVx}
\end{figure}

In Figure \ref{fig:HighLowVx} and movies LowerVx.avi and HigherVx.avi we compare the evolution of the members of the ensemble generated with the lower ($|v_x|\sim  363$ km/s, panel (a)) and higher ($|v_x| \sim $583 km/s, panel (b)) absolute value of the $v_x$ velocity component. The movies show that the velocity values and magnetic field line patterns are significantly different from the reference simulation and from the ensemble average, demonstrating that the perturbations are not trivial.

\begin{figure}[h!]
\begin{center}
\includegraphics[width=0.9\textwidth]{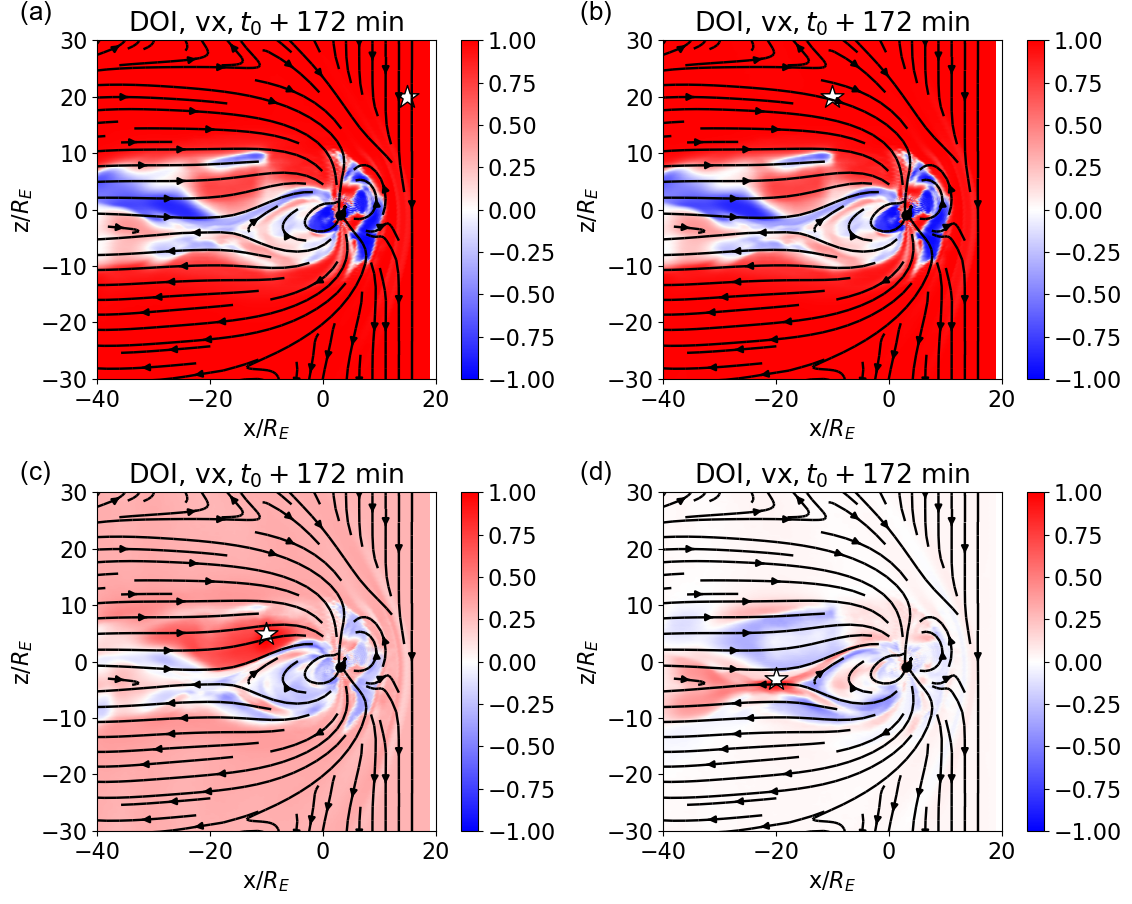}
\end{center}
\caption{DOI maps for $v_x$, computed from the correlation of an ensemble of OpenGGCM magnetospheric simulation, with observation points in the solar wind (panel (a)), magnetosheath (panel (b)), northern lobe (panel (c)), plasma sheet (panel (d)).}
\label{fig:DOI_vx}
\end{figure}

\begin{figure}[h!]
\begin{center}
\includegraphics[width=0.9\textwidth]{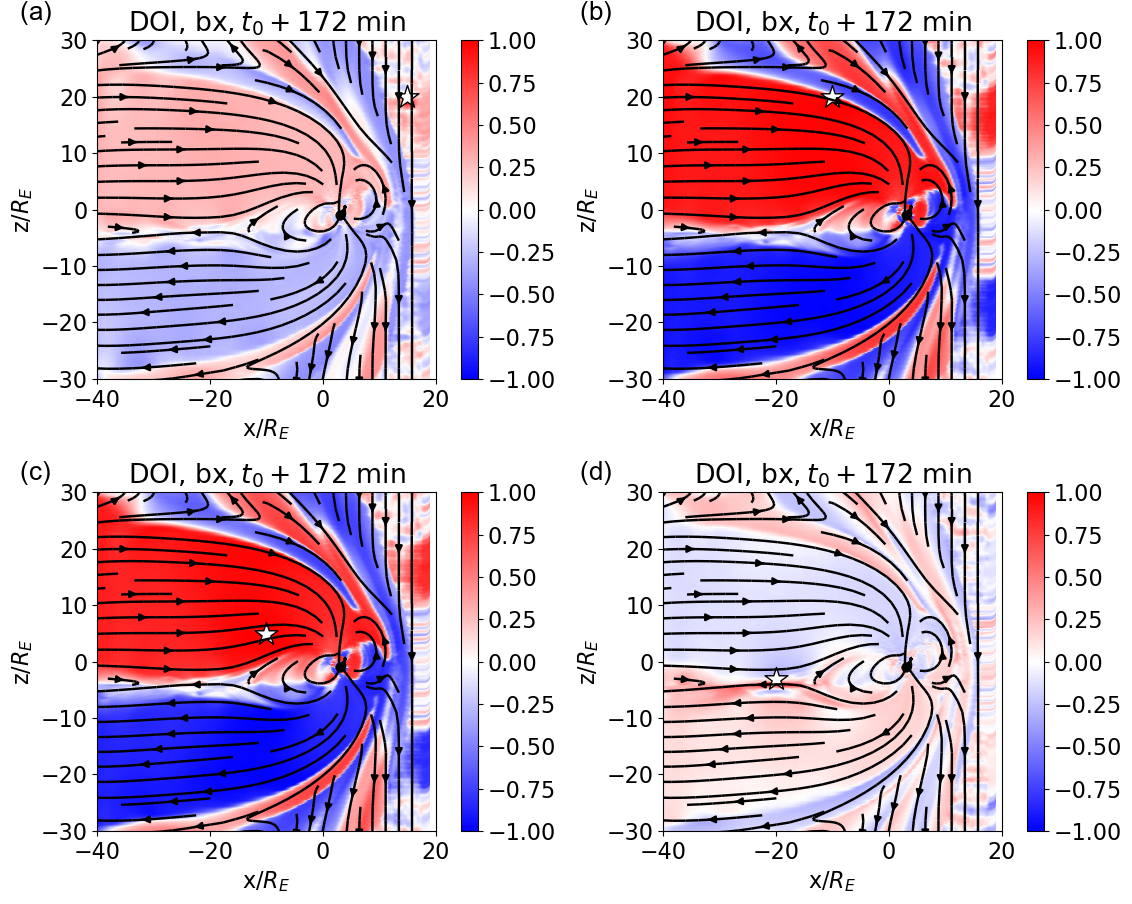}
\end{center}
\caption{DOI maps for $b_x$, computed from the correlation of an ensemble of OpenGGCM magnetospheric simulations, with observation points in the solar wind (panel (a)), magnetosheath (panel (b)), northern lobe (panel (c)), plasma sheet (panel (d)).}
\label{fig:DOI_bx}
\end{figure}

We now discuss the RA and DOI analysis for a set of different observation points, depicted as white stars in the following figures, in the inflowing solar wind (a), in the magnetosheath (b), in the northern lobe (c),  and in the plasma sheet (d), for the same plane and time as Figure~\ref{fig:OpenGGCM_1}. The coordinates of each of these points in the $xz$-plane are given in Table~\ref{tab:coords}, the $y$ coordinate being $y/R_E=0$. Figure~\ref{fig:DOI_vx} and Figure~\ref{fig:DOI_bx} show the DOI maps for $v_x$ and $b_x$ respectively. Note that the correlations which are displayed are not cross-correlations: the correlation is done between the value of a field at the observation point and the values of the same field at the other grid points.

\begin{table}[]
\begin{center}
\begin{tabular}{l|ccc}
 \toprule
 & $x/R_E$ & $y/R_E $ &$ z/R_E$    \\
  \hline
solar wind & 15 &0 & 20   \\
magnetosheath & -10 &0 & 20  \\
northern lobe & -10 &0 & 5   \\
plasma sheet &-20  & 0 &  -3 \\
 \bottomrule
\end{tabular}
\caption{Coordinates of the observation points used in the DOI analysis. 
}
\label{tab:coords}
\end{center}
\end{table}

Figures~\ref{fig:DOI_vx} and~\ref{fig:DOI_bx} show that the correlations are mostly ordered by the principal regions of the magnetosphere such as the lobes, the  magnetosheath, and the plasma sheet.  For example, Figure~\ref{fig:DOI_vx} shows the results for the  $v_x$ correlations. The DOI values for the plasma sheet are different from those in the magnetosheath and the lobes in all panels. As expected, the stronger correlations are somewhat localized around the observation point, for example, the strongest correlations in panel (d), where the observation point is in the plasma sheet, are in the plasma sheet itself and its immediate surroundings.  However, some other observation points have a much larger DOI, such as the ones in the solar wind and the magnetosheath.  This makes physical sense, because $v_x$ variations in those regions will propagate through much of the magnetosphere.  Figure~\ref{fig:DOI_bx} shows the $B_x$ correlations. The northern and southern lobes clearly stand out, with opposite DOI values, and the magnetosheath stands out as well. Because the $B_x$ values have opposite signs in the two lobes, the DOI values also have opposite signs.  The correlation values depend on the variability of the field value at the observation point, thus the panels that exhibit lower correlations are those with observation point in the plasma sheet, where the intrinsic variance of both $v_x$ and $B_x$ is higher (see Figure~\ref{fig:OpenGGCM_1}) due to the different reconnection patterns in the different members of the ensemble. For example, in Figure~\ref{fig:DOI_vx}, panel (d) the velocity value at the observation point in the plasma sheet exhibits little  correlation with the $v_x$ values outside of the plasma sheet and the neighboring areas. This is a consequence of the jet structure which is caused by internal magnetospheric dynamics rather than the solar wind driver.

The temporal dynamic DOI behavior is similar: the DOI maps of $v_x$ and $B_x$ with observation point in the plasma sheet exhibit higher temporal variability than those with a observation point in the magnetosheath, {as can be seen in the DOI movies DOI\_bx\_bx\_MSheath.avi, DOI\_bx\_bx\_pSheet.avi, DOI\_vx\_vx\_MSheath.avi, DOI\_vx\_vx\_pSheet.avi and in Figure~\ref{fig:OtherTIME}. The Figure shows the DOI map for $v_x$ (panel (a) and (b)) and $B_x$ (panel (c) and (d)) with observation point in the plasma sheet (panel (a) and (c)) and in the magnetosheath (panel (b) and (d)) at $t_0$ + 192 minutes. All the previous figures,  Fig.~\ref{fig:OpenGGCM_1}, ~\ref{fig:HighLowVx}, ~\ref{fig:DOI_vx} and~\ref{fig:DOI_bx}, were at $t_0$ + 172 minutes. 

We note that the plots with observation points in the magnetosheath are not significantly different to earlier plots (see Figures~\ref{fig:DOI_vx}, \ref{fig:DOI_bx}), except for the plasma sheet plots, which differ profoundly.

To summarize and interpret the OpenGGCM results, the DOI analysis is well in line with our understanding of the terrestrial magnetosphere. In the $v_x$ case, when the observation point is in the solar wind or in the magnetosheath, the $|DOI|$ values are very high in both the solar wind and the magnetosheath region. This is expected, because $v_x$ in the solar wind is a correlation with itself (and thus a sanity test for the calculation), whereas the magnetosheath is largely driven by the interaction between solar wind and the bow shock, where the Rankine-Hugoniot conditions predict a positive correlation of the downstream velocity with the upstream velocity. When the observation points are in the solar wind and magnetosheath regions, the $|DOI|$ values in the plasma sheet are expected to be lower due to internal transient dynamics (e.g., reconnection events, bursty bulk flows) in the sheet which may be triggered by local plasma sheet dynamics, rather than solar wind compression. Local dynamics in the sheet is also the reason why, in Figure~\ref{fig:DOI_vx}, panel (d), when the observation point is in the plasma sheet, the correlation with the solar wind and magnetosheath regions is close to zero. Even if, in global terms, magnetic reconnection in the plasma sheet were triggered by magnetopause dynamics, in any region of the plasma sheet $v_x$ may flow sunward or anti-sunward, depending on the location of the reconnection site, and thus would be uncorrelated with the velocity in the solar wind or in the magnetosheath. The lobe magnetic field is expected to be directly driven by the solar wind dynamic pressure, and thus by $v_x$.  As the dynamic pressure increases, the lobe flare angle decreases, and vice versa.  As the flare angle decreases, the lobe field gets compressed. Figure \ref{fig:DOI_bx}, panels (b) and (c) show that effect, as expected.

Similar consideration broadly apply to the $B_x$ DOI results shown in Figure~\ref{fig:DOI_bx}. There is, however, a significant difference between panels (a) in~\ref{fig:DOI_vx} and~\ref{fig:DOI_bx}. When the observation point is in the solar wind, high correlations are obtained in large parts of the magnetosphere for $v_x$, while $B_x$ correlations are much lower. This can be attributed to the fact that $B_x$ in the solar wind is not a major driver of magnetospheric dynamics, unlike the solar wind speed and solar wind dynamic pressure. The geo-effective component of the interplanetary magnetic field (IMF) is the $B_z$ component, which controls reconnection at the magnetopause and thus the dominant energy input into the magnetosphere.

\begin{figure}[h!]
\begin{center}
\includegraphics[width=0.9\textwidth]{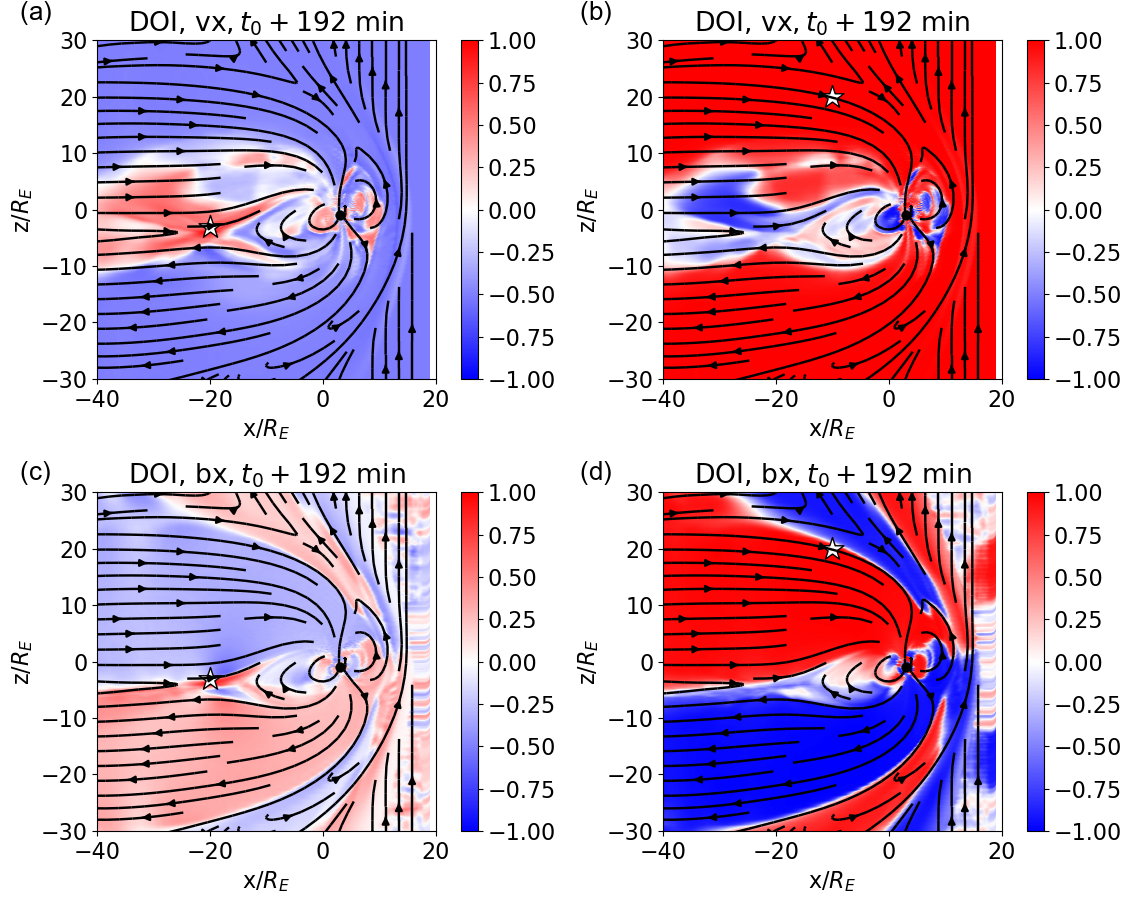}
\end{center}
\caption{DOI maps for $v_x$ (panel (a) and (b)) and $B_x$ (panel (c) and (d)) from an ensemble of OpenGGCM magnetospheric simulations with observation point in the plasma sheet (panel (a) and (c)) and in the magnetosheath (panel (b) and (d)), at $t_0$+ 192 minutes. All previous Figures were at  $t_0$+ 172 minutes.}
\label{fig:OtherTIME}
\end{figure}

\subsection{Magnetospheric applications II: Tsyganenko Model}
\label{sect:Tsyg}

The Tsyganenko models are a family of empirical, static terrestrial magnetic field models~\citep{tsyganenko1987global, tsyganenko1989magnetospheric,tsyganenko1995modeling, tsyganenko2002modela, tsyganenko2002modelb, tsyganenko2003storm, tsyganenko2005modeling}. The successive model versions (available at~\url{http://geo.phys.spbu.ru/~tsyganenko/modeling.html}) reflect increasing knowledge of the magnetospheric systems and are based on an increasing amount of data from all regions in the magnetosphere. 

The models are based on a mathematical description of the magnetosphere, which includes contributions from major magnetospheric current sources such as the Chapman-Ferraro current, the ring current, the cross-tail current sheet and large-scale field-aligned currents. Terms are added to account for the magnetopause and for partial penetration of the IMF into the magnetosphere. The most recent versions can also take into account the dipole tilt, the dawn-dusk asymmetry, and allow for open magnetospheric configurations. The parameters of the models are derived from a regression to magnetic field observations, and keyed to magnetic indices  and/or solar wind parameters. The model requires the user to specify a date and time for the dipole orientation. The other model parameters, either an index such as the Kp, or solar wind variables, are to be given by the user. In more recent models, Tsyganenko also provides yearly input data files for his models. From these inputs, an approximation of the magnetosphere is created for the specified date and time. Notice that the Tsyganenko models are static, and only provides a snapshot of the magnetosphere. However, since the parameters are time dependent the model can be used in a quasi-dynamic mode. 

Several versions of the Tsyganenko model have been tested over the years against observations and physics-based, MHD models~\citep{thomsen1996observational, huang2006storm, woodfield2007comparison}. While the Tsyganenko models do not account for the Earth's internal magnetic field, methods are provided to add the internal field model as described in the above cited literature.

In order to simulate the evolution of the magnetosphere with the chosen Tsyganenko model, we create snapshots of the magnetosphere at different times. The time May 8\textsuperscript{th}, 2004, 09:00 UTC is taken as $t_0$, the same time as the OpenGGCM simulations presented in Section \ref{sect:OpenGGCM}. The model is ``evolved" by using a time series of the required input parameters,  which are obtained from the OMNIWeb database~\citep{king2005solar}. 

We use two versions of the Tsyganenko model, the T96 model \citep{tsyganenko1996effects} and the TA15 model \citep{tsyganenko2015forecasting}. We generate the Tsyganenko ensembles in the same way as the OpenGGCM examples, by using a distribution of $v_x$ values as described in section~\ref{sect:OpenGGCM}.  

Before we analyse the results of the T96 ensemble, we show the magnetospheric configuration computed by the model using the original solar wind data. In Figure \ref{fig:T96_ref}, the first row of figures shows the results of the ``reference" simulation, e.g. the simulation without any perturbed inputs, at time $t_0 + 85$min, for $B_x$ (panel (a)) and $B_z$ (panel (b)). 

Unlike the OpenGGCM, the T96 model cannot model reconnection, although some approximation of reconnection is included in later Tsyganenko models \citep{tsyganenko2002modela, tsyganenko2002modelb}. Also, the day-side magnetospheric structure is only approximated with respect to physics-based models, and bow shock and magnetosheath are not clearly distinguishable. In Figure~\ref{fig:T96_ref}, the second row shows the average of the ensemble at the same time of the reference simulation, for $B_x$ (panel (c)) and $B_z$ (panel (d)). The results are similar to the reference simulation, as shown by the logarithm of the absolute difference between the reference and ensemble mean, e.g., Figure \ref{fig:T96_ref}, panel (e) and (f). The only significant difference is located at the magnetopause, which is expected since varying the solar wind velocity changes the standoff distance.

\begin{figure}
    \centering
    \includegraphics[width=.8\textwidth]{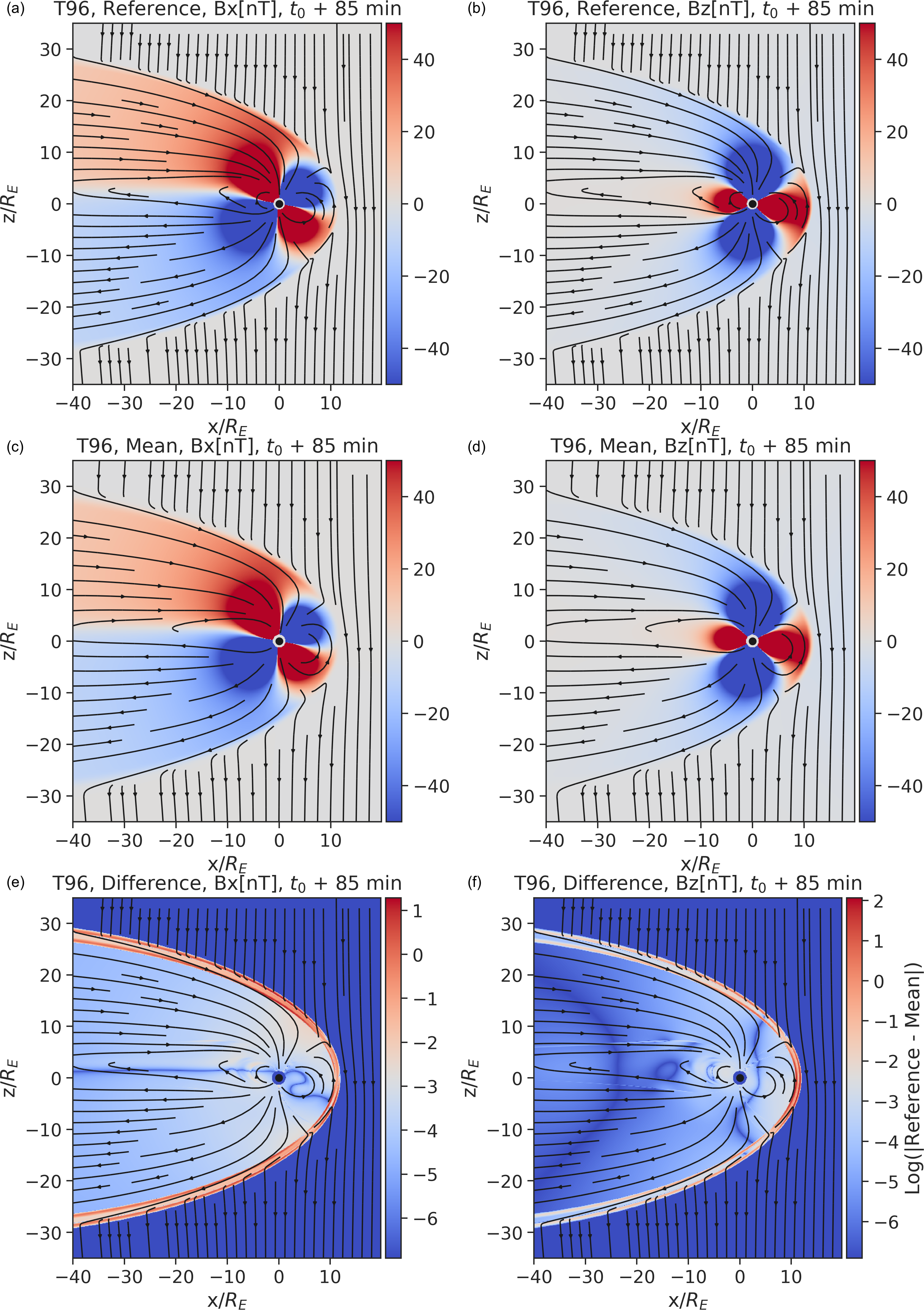}
    \caption{Results from the T96 Tsyganenko model. The top row shows the reference magnetic field ($B_x$ and $B_z$ component) at time $t_0+ 85$ min, with a negative IMF. The middle row shows the ensemble mean of the same magnetic field components at the same time $t_0+85$min. We clipped the magnetic field values to $|50 nT|$, in order to make variations in the tail better visible. The last row shows the logarithm of the absolute difference between the reference simulation and the mean of the ensemble.}
    \label{fig:T96_ref}
\end{figure}

Next we analyse the DOI maps of the T96 model. Figure \ref{fig:T96_85} shows the DOI maps for the $B_x$ and $B_z$ field components at time $t_0+85$ min. Although the T96 model is parameterized by the solar wind velocity, it only models the magnetic field in the magnetosphere. Because of this, we are only able to analyze the DOI maps of the magnetic field components. The observation points are placed in the northern lobe, in proximity to the current sheet, at the dayside magnetosphere, and in the southern lobe. Like in the OpenGGCM case, the DOI maps reflect the general regions of the magnetosphere as reproduced by the T96 model. However, the correlation only takes values of $\pm$1 in the magnetosphere, and zero in the solar wind.  The latter is simply a consequence of the fact that the model does not predict the IMF, which is therefore independent of the $v_x$ variations of the ensemble.
The former is due to the fact that the model has no intrinsic time dependence.  Any variations of the solar wind affect the entire magnetosphere instantly and in proportion to the variation.  Thus, after normalization, only the sign matters, i.e., whether a given change at the observation point leads to a positive or negative change at a different point. 

\begin{figure}
    \centering
    \includegraphics[width=.8\textwidth]{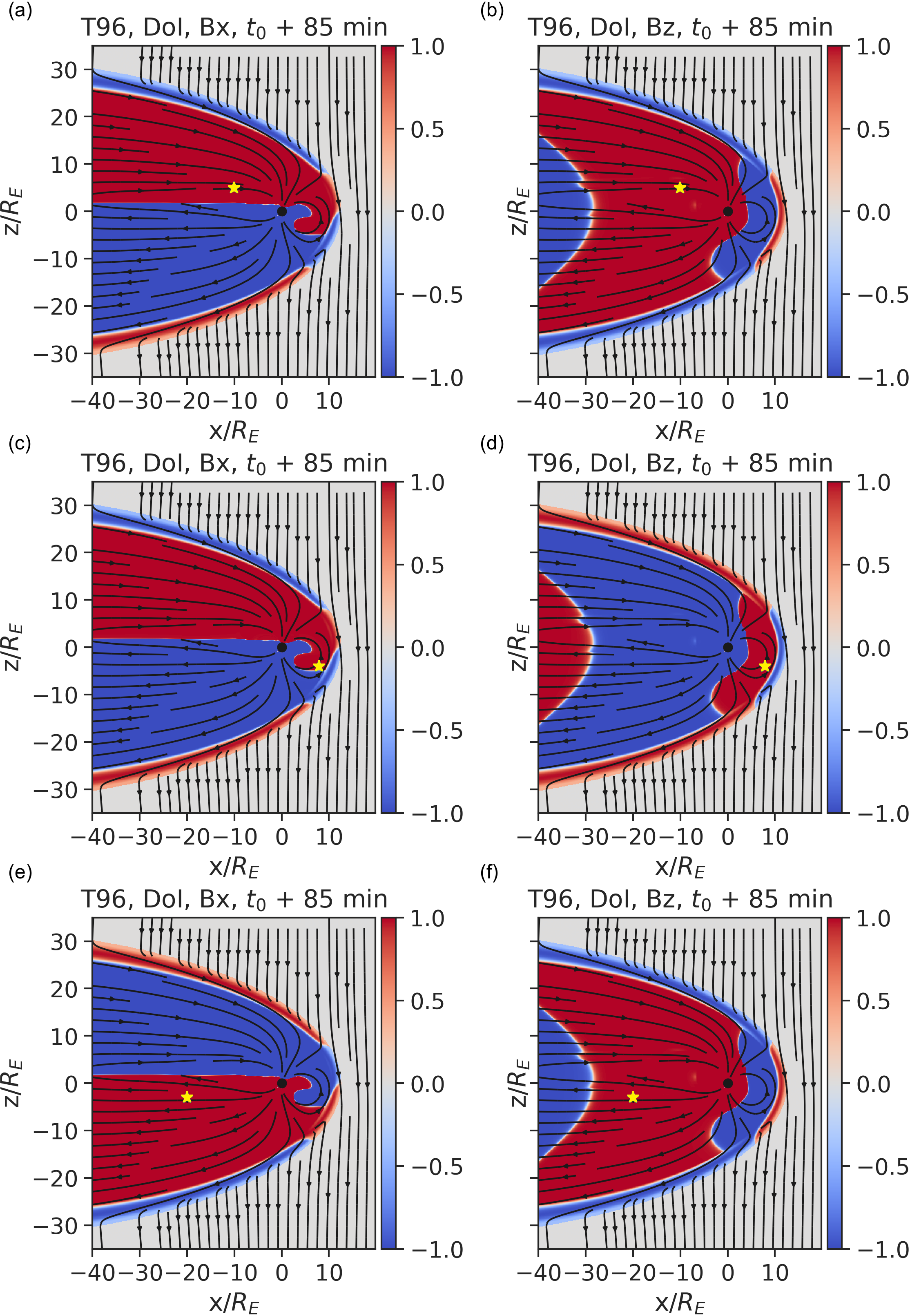}
    \caption{Each row displays the DOI maps for $B_x$ and $B_z$, from an ensemble of Tsyganenko model T96 simulations with observation point in the plasma sheet, dayside magnetosphere and southern lobe tail, respectively.}
    \label{fig:T96_85}
\end{figure}

Now we focus on the results of the TA15 model. Figure~\ref{fig:TA15_mean_ref_85} shows the reference simulation and ensemble mean for the $B_x$ and $B_z$ fields, with superimposed field lines, at time $t_0+85$ min, together with the difference between reference and ensemble mean. We observe that the reproduced dayside magnetosphere structure is improved compared to the T96 model, at the expense of unrealistically high magnetic field values in the inflowing solar wind, and correspondingly distorted magnetic field lines. These artificial boundary conditions in the Sunwards boundary are used to obtain an ``open" magnetosphere which blends with the inflowing solar wind, without seeming to form a nightside magnetosheath. Notice also the high values at these artificial boundary conditions in the difference plot, indicating that there is a high variability in their values.

From the DOI maps in Figure~\ref{fig:TA15_DOI} (with observation points at the same positions as Figure~\ref{fig:T96_85}), we can confirm that the modelled IMF is used to construct the internal magnetospheric solution. While in the T96 model the solar wind $B_x$ and $B_z$ values were uncorrelated with the magnetospheric values, here the absolute value (i.e. ignoring the sign) of the correlation is very high: the solar wind input strictly determines the inner magnetospheric solution, making the correlation practically unitary. This could be because of the deterministic analytical formula used to construct the magnetic field, where everything is exactly determined on a global scale.

Note that the correlations reported are spatial and not temporal, therefore no causality is implied. High correlation between the IMF and magnetospheric fields point to the fact that, in an ensemble generated by perturbing the solar wind input, the model is built in such a way that variations in the magnetic field are highly correlated through the system, apparently without highlighting the boundary regions that we were able to spot in the DOI maps for the OpenGGCM and Tsyganenko T96 simulations. 

\begin{figure}
    \centering
        \includegraphics[width=.8\textwidth]{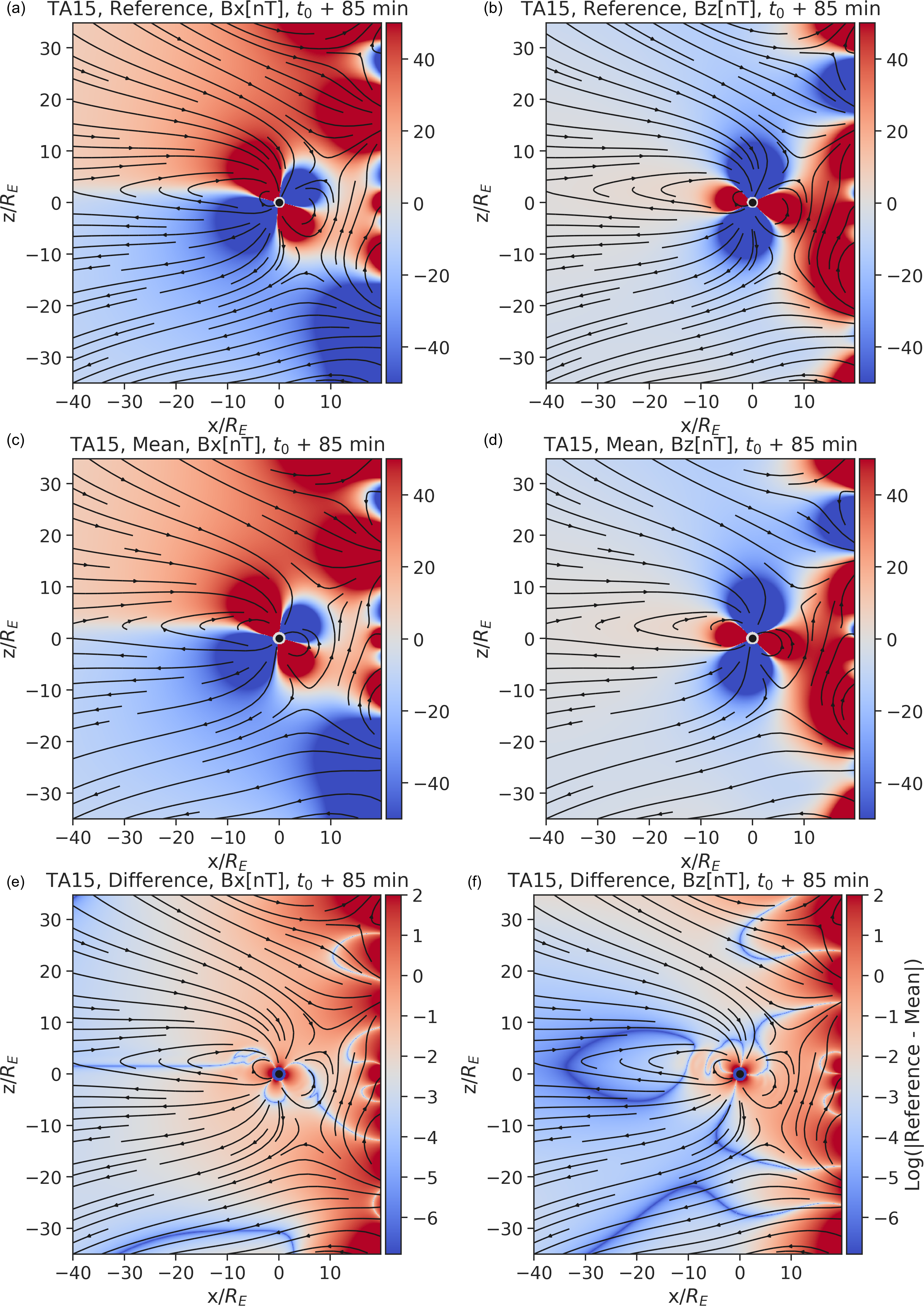}
    \caption{The reference and mean magnetic field at time $t_0+85$ min from the Tsyganenko TA15 model are displayed in the top two rows of figures. Notice the non-realistic high magnetic field values in the inflowing solar wind (at $x/R_E > 10$). We assume these unrealistic values are necessary for the model to construct the day-side magnetosphere.The values of the magnetic field have been cut off in the top two rows of figures at $|50 nT|$, to make sure the variations in the tail are visible. The last row of figures shows the logarithm of the absolute error between the reference simulation and the mean of the ensemble.}
    
    \label{fig:TA15_mean_ref_85}
\end{figure}

\begin{figure}
    \centering
    \includegraphics[width=.8\textwidth]{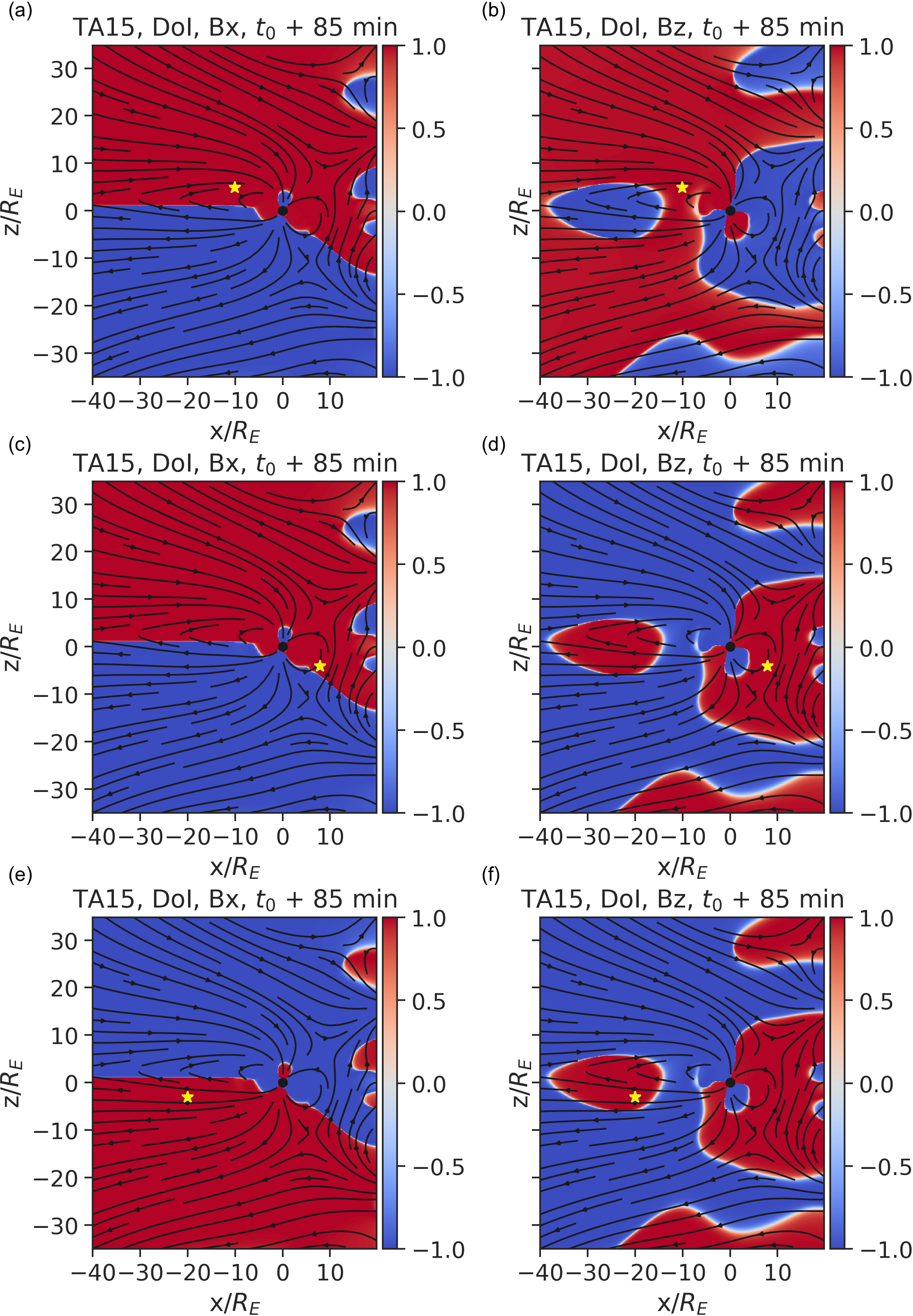}
    \caption{DOI maps for $B_x$ and $B_z$, computed from an ensembles of Tsyganenko TA15 simulations at $t_0+85$ min with observation points at the same position as Figure~\ref{fig:T96_85}.}
    \label{fig:TA15_DOI}
\end{figure}

A last remark on the DOI analysis applied to the Tsyganenko models is the following. The analysis helps us understand and visualize how the different models are built, with regards to the relationship between the solar input and the magnetospheric solution. DA analysis then proves useful here as a model investigation tool. 

It also highlights that caution should be used when deciding to apply DA techniques to a particular model, depending on the objectives of the investigation. The Tsyganenko models were built to provide time-independent, empirical-based insights into the structure of the magnetosphere at a particular instant in time. They do not aim at representing the state of the pristine solar wind, which is used only to better the magnetospheric solution (hence the somehow unrealistic solar wind patterns identified in Figure~\ref{fig:TA15_mean_ref_85} and~\ref{fig:TA15_DOI}). Also, they do not intend to reproduce temporal dynamics in the magnetosphere. These factors result in DOI maps where the absolute value of the correlation is always either 1 or -1. When using DOI techniques with the purposes of identifying useful locations for satellite placement, these are not useful results: we are interested in the \textit{value} of the DOI, not in the \textit{sign}. Hence, caution should be used before using empirical, time-independent models for this particular purpose: more significant information will possibly be acquired from their physics-based, time-dependent counterparts. This consideration does not intend to diminish the importance of empirical, time-independent models for other scientific objectives such us, most importantly, quick forecasting. }

\subsection{Heliospheric application: PLUTO} \label{sect:PLUTO}

In this section, we study the propagation of a Coronal Mass Ejection in a solar meridional plane, which is defined by the rotation axis of the Sun and a radial vector in the equatorial plane. In all the runs of the ensemble, the computational domain is $1R_{\odot} \leq r \leq 216 R_{\odot}$ and $0 \leq \theta \leq \pi$ in spherical coordinates, where $R_{\odot}$ is the solar radius and $\theta$ is the polar angle (or colatitude), corotating with the Sun. Assuming axisymmetry around the solar rotation axis, we may limit our analysis to 2.5D (pseudo 3D) simulations. The grid resolution is uniform in both directions, $384\times384$ cells, which is sufficient to capture the structure of the background solar wind while keeping the computational cost and output size manageable. 

We simulate the background solar wind using a simple adiabatic model with effective polytropic index $\Gamma=1.13$ \citep{Keppens2000}. We also assume a time-independent dipole background magnetic field:
\begin{center}
\begin{align}
B_r &= - 2 B_o \cos \theta / r^3, \\
B_{\theta} &= - B_o \sin \theta / r^3,
\end{align}
\end{center}
\noindent
where $B_r, B_{\theta}$ are the $r, \theta$ components of the magnetic field in spherical coordinates and $B_o$ a constant used to scale the field to $B=1.1$G on the solar equator. We impose the density distribution $\rho$ as a function of the latitude $\theta$ at the inner boundary to achieve a “dead zone” of low velocity near the equator and a fast solar wind near the poles simultaneously (see \cite{Keppens2000} and \cite{Chane2008}). The differential rotation of the Sun is also taken into account, following \cite{Chane2005}; this is achieved by imposing a varying azimuthal velocity $v_{\phi}=v_{\phi}(\theta)$ at the inflow boundary.

Once the simulation reaches a steady state, roughly after $\sim~2.5$ days or $t=10$ in normalized units, the radial velocity at 1 AU is $\sim 300$ km/s near the equator and $\sim 850$ km/s near the poles. This is consistent with the large-scale bimodal solar wind structure that is typically observed during solar minimum~\citep{McComas1998}).

\begin{figure}
\centering
\includegraphics[scale=0.65]{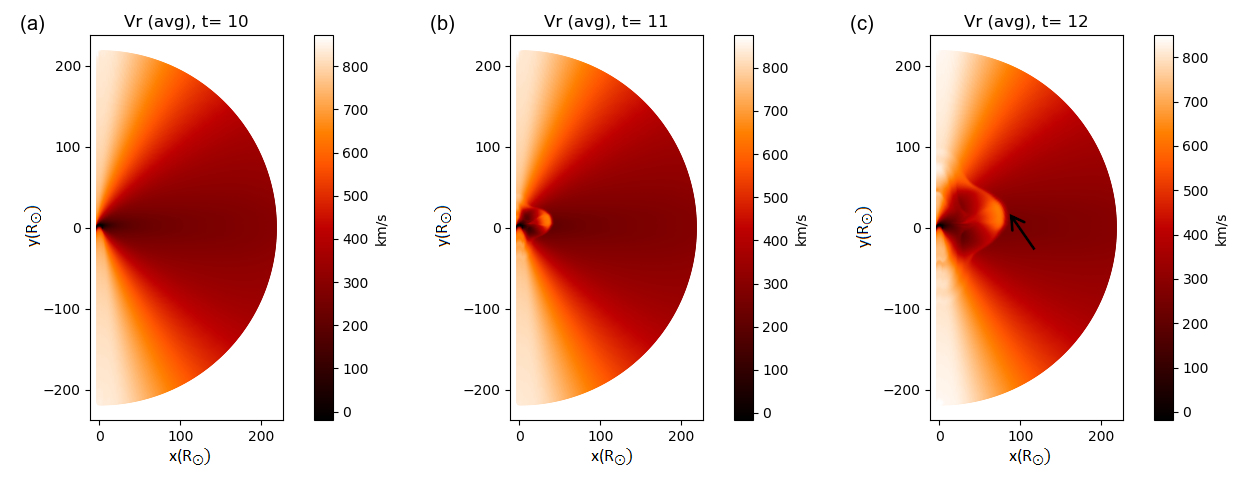}
\caption{Injection and propagation of a CME via the average radial velocity in the ensemble (in km/s). From left (a) to right (c): relaxed solar wind (t=10), injection and propagation of the CME ($t=11$, $t=12$). The leading edge (front shock) is prominent in the last panel, marked by a black arrow.}\label{fig: windcme}
\end{figure}

We create two ensembles of 100 simulations each. In the first ensemble, the velocity of the CME in each case is randomly selected from a Gaussian distribution with mean $\mu=900$ km/s and standard deviation $\sigma=25$ km/s. The resulting values are typical of strong CME events. In the second ensemble, the spatial extent of the boundary conditions that launch the CME varies as well, along with the velocity of the CME as described above. The half-width of this region is also randomly selected from a Gaussian  distribution with $\mu=10^{\circ}$ and $\sigma=0.5^{\circ}$. All other parameters remain the same in every run. 

The values of the CME widths that are used here are comparable to observed events. The choice of parameters in the second ensemble is less constrained by observations and leads to the appearance of very small values of variance. We  thus find large areas where the DOI $\sim 1$, since the simulations in the ensemble do not differ significantly. This was confirmed by creating and analyzing a third ensemble, where the width of the CME is chosen from a Gaussian with $\mu=20^{\circ}$ and $\sigma=2.0^{\circ}$. 

Figure~\ref{fig: windcme} shows the evolution of the radial velocity average over the whole ensemble. Up to t=11, i.e., before the CME is launched, all runs are identical.   The CME is initialized similar to the simplified approach of \cite{Keppens2000}, such that the boundary conditions on the solar surface are modified to represent a change of mass flux. In our case, we modify the boundary conditions at R=1R$_\odot$, in a given region around $\theta=80^{\circ}$. A tracer (a passive scalar only present as an advected quantity within the flow, without effect on the plasma) is also injected with the CME, to facilitate monitoring its propagation. In the middle and right panel of Fig.~\ref{fig: windcme} we show the ejection of the CME and its propagation. The CME front can be clearly distinguished at t=12.

To apply the RA technique, as explained in Section \ref{sec: Methods}, we select a point of interest and perform the analysis based on (a) the plasma density, or (b) the radial velocity.  We present results at t=14, when the CME has reached a distance of $\sim 150R_{\odot}$, for two detection points (at $R=90$ and $R=150R_{\odot}$, $\theta=80^{\circ}$). At times earlier than $t=10$ (when the solar wind reaches a steady state and the CME is injected), the DOI is zero, since the observation point is disconnected from the rest of the domain before the CME reaches it.

The propagation of the CME can be monitored in the MHD simulations easily via e.g. a tracer (or the radial velocity). The DOI map, when the tracer is used as a criterion, follows closely the CME propagation pattern observed in the MHD runs. However, this is of limited use, besides testing, as the tracer (in our case) does not represent a real physical quantity. 

\begin{figure}
\centering
\includegraphics[width=0.875\textwidth]{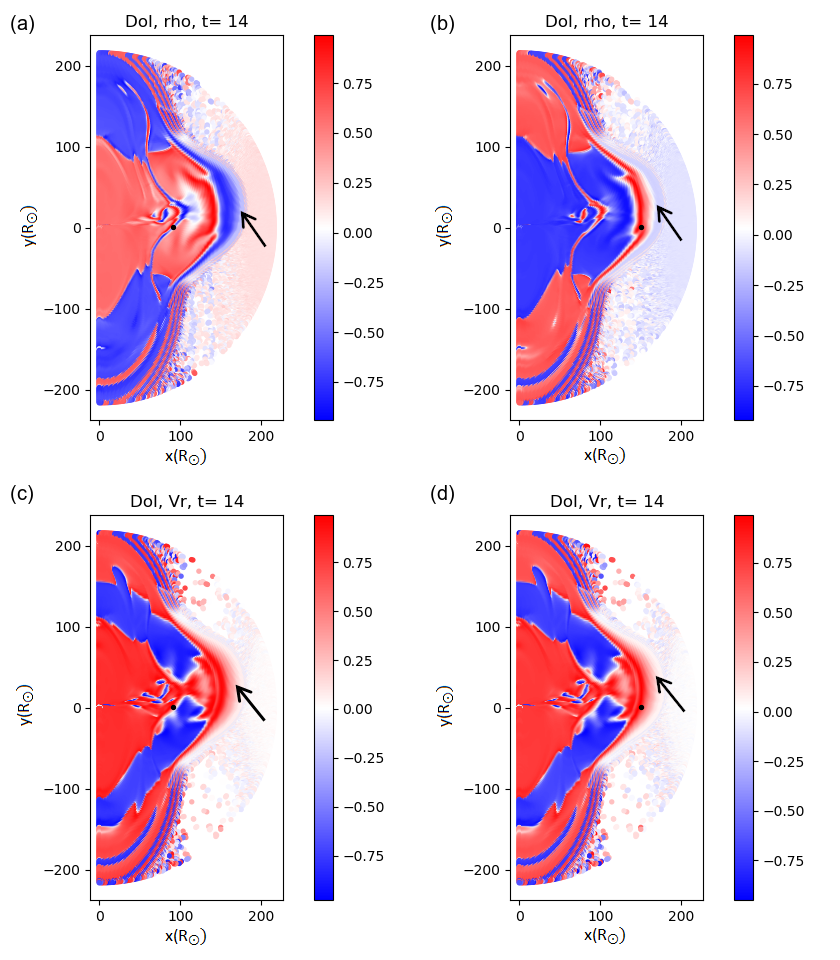}
\caption{DOI maps for the density (first row) and the radial velocity (second row), from an ensemble of PLUTO simulations. The first column has the observation point at $R=90R_{\odot}$ and the second column at $R=150R_{\odot}$ (marked by the black dot). In this set we only perturb the radial velocity of the CME. The structure of the CME can be seen quite clearly in panels (a) and (b), where the leading edge is evident and marked by a black arrow in all panels.}\label{fig:doi1v}
\end{figure}

The DOI map for the first ensemble, where we perturb only the radial velocity of the CME, is shown in Fig.~\ref{fig:doi1v}. The regions where information from the CME front has not yet arrived have DOI=0, as shown in the radial velocity DOI map (Fig.~\ref{fig:doi1v}). When only one parameter is modified (first ensemble), the density DOI map shows a very large area of the domain saturated with correlation$\simeq$1. This is probably due to variations in density of the background solar wind induced by the propagation of the CME. The regions with absolute value of the DOI$\simeq1$ that are located far from the CME propagation front (at small or large angles $\theta$) are the areas of high radial velocity in Fig.~\ref{fig: windcme}, where the information on the perturbation introduced when triggering the CME has already propagated. The density and radial velocity of all ensemble members are modified in a similar way, hence the large $|DOI|$ values. 

\begin{figure}
\centering
\includegraphics[scale=0.875]{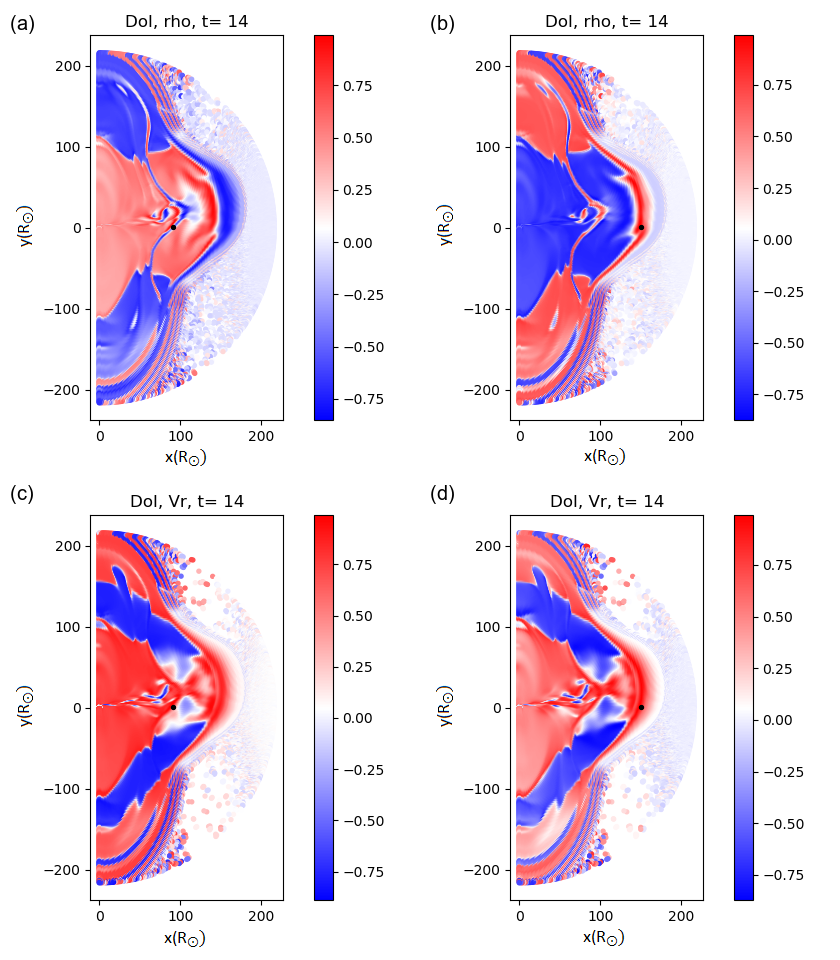}
\caption{DOI maps for the density (first row) and the radial velocity (second row), computed from PLUTO simulation ensembles. The first column has a detection point at $R=90R_{\odot}$ and the second column at $R=150R_{\odot}$ (marked by the black dot). In this set we perturb two parameters, the radial velocity and the size of the CME.}\label{fig:doi2v}
\end{figure}

In the second ensemble, where the width of the CME is also modified, the DOI map of the density shows smaller correlation values (compare especially panel (b) and (d) in the two Figures) compared to the previous ensemble, because the differences between the runs of the ensemble are now larger (see Fig.~\ref{fig:doi2v}). This results in smaller regions where the DOI is close to unity compared to the first case. 

\begin{figure}
\centering
\includegraphics[scale=0.875]{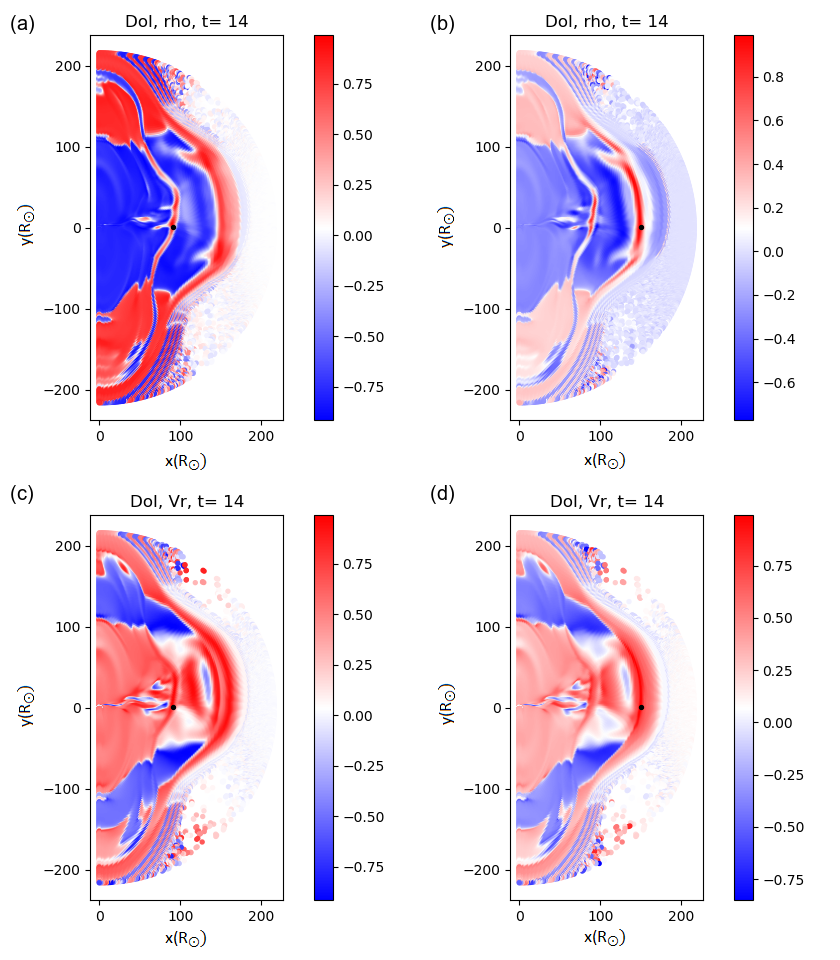}
\caption{DOI maps from ensembles of PLUTO simulations for the density (first row) and the radial velocity (second row). The first column has a detection point at $R=90R_{\odot}$ and the second column at $R=150R_{\odot}$. In this set we perturb two parameters, the radial velocity and the size of the CME, with a larger perturbation in the size with respect to the second data set.}\label{fig:doi3v}
\end{figure}

The last ensemble, where the CME width and its perturbation are larger compared to the second case, is shown in Fig.~\ref{fig:doi3v}. The DOI pattern is qualitatively similar to Fig.~\ref{fig:doi2v}, but due to the larger values in the size of the CME and its perturbation, the regions with high DOI values (meaning the regions affected by CME propagation in at least one member of the ensemble) are slightly larger as well. 

Additional analysis, not shown here, was carried out on subsets of the ensembles to ensure the ensemble size is sufficient. We found that in this case convergence was achieved if at least 25 ensemble members were used (as described in Section~\ref{sec: Methods}); however, this number may differ in other cases, depending on the specifics of the ensemble.

The DOI analysis applied to the simulations performed with PLUTO are indicative of the versatility of the method. In all PLUTO ensembles, we can monitor the influence of the CME during its propagation and the response of the system via the DOI. Moreover, we can identify certain CME components, such as the leading edge, from the DOI maps. Differences in the response of the system due to the choice of the perturbation or parameters are captured as well. The resolution used here was sufficient to capture the CME injection and propagation within reasonable computational cost; the typical run time for simulating a member of the ensemble was of the order of $\sim10'$ minutes on 28 cores.

However, some limitations of the model must be considered. The \textbf{axisymmetric} assumption simplifies the problem and allows to reduce computation costs, but with the drawback of not accounting for the three-dimensional CME structure. The limited angular resolution imposes a weak constraint on both the perturbed and unperturbed size of the CME \textbf{that we can simulate}. Runs with a higher resolution can remove this constraint at additional computational cost. Simulations in 3D will be part of future work in order to capture the full system, where also differences in the polar direction can be examined. Finally, a more realistic model for the background magnetic field should be used, rather than a simple static dipole. We focused mainly on calculating the DOI at different times and locations, but a similar approach can be used to estimate the arrival time of the CME, as described in \cite{Owens2020}.

\FloatBarrier 

\section{Summary and Conclusions} \label{sec: Conclusions}
In this paper, we apply the Representer analysis and the Domain of Influence analysis to two fundamental components of the Sun-Earth connection: the interaction between the solar wind and the terrestrial magnetosphere, simulated with the OpenGGCM MHD code and with the empirical Tsyganenko models, and the propagation of CMEs in the background solar wind, simulated with the MHD PLUTO model.

In each case an ensemble is generated by appropriately perturbing initial/ boundary conditions. Subsequently, the DOI analysis is applied over the ensemble. Localisation methods, which can be used to reduce spurious correlations in the estimated prior covariance matrix \citep{anderson2007exploring, bishop2007flow, sakov2011relation}, are not used at this stage.

Primarily, the DOI analysis is a first step in the application of Data Assimilation techniques to a model, and can be applied before assimilation itself to gain insight on the system and on the model. However, the DOI analysis can also be used to gain physical insight, and to devise optimized observation systems, as discussed below.

Our main results are as follows.

First, we have demonstrated that DOI analysis can provide useful information on the most appropriate locations for future observation points, such as solar wind and magnetospheric monitors. Large absolute values of the DOI, calculated with respect to an observation point, means that observations at that location would provide significant information of that field in the specific, large $|DOI|$ area, but less so in areas with lower $|DOI|$. This can be used in two different ways.
On one hand, DOI analysis can help to find observations points that are connected to large $|DOI|$ areas, in order to increase the amount of information brought in by a single new observation.
On the other hand, the same information can be used for a different objective. Given a particular location, one can ask where observations need to be obtained to improve knowledge of that area. A useful example here is the plasma sheet in the OpenGGCM analysis, Section~\ref{sect:OpenGGCM}. Figure~\ref{fig:DOI_vx} and~\ref{fig:DOI_bx} show that $|DOI|$ values in the plasma sheet are consistently low, notwithstanding the field which is examined ($v_x$ or $B_x$) and the location of the observation point. $|DOI|$ values in the plasma sheet are low even if the observation point is in the plasma sheet itself: $|DOI|$ values, which are of course 1 at the observation point itself, quickly become smaller even a small distance away. Since the plasma sheet is a location of particular importance for space weather forecasting, or basic research for that matter, single s/c in the plasma sheet are of limited use, and rather a constellation of satellites, such as proposed in~\citet{angelopoulos98a, raeder98e} would be necessary.

Second, we have used the DOI analysis to improve our knowledge of the models we use, and in particular to investigate whether these models are appropriate for the implementation of Data Assimilation. The DOI analysis for the Tsyganenko models in Section~\ref{sect:Tsyg} powerfully highlights the model evolution from version T96 to version TA15. In version T96, the magnetosphere is a closed system, and solar wind conditions are not correlated (DOI $\sim$ 0 in Figure~\ref{fig:T96_85}) with the magnetospheric region. In version TA15, the magnetosphere opens up to solar wind driving, and the correlation between the solar wind region and the magnetosphere becomes very high (Figure~\ref{fig:TA15_DOI}). One should remember that the Tsyganenko models are supposed to be used to investigate the magnetospheric system, and the solar wind configuration is artfully modified as to give the best representation of the magnetosphere under the specified conditions.  

One common aspect of the two versions of the Tsyganenko models is that (with the exception of the solar wind region in T96) the DOI values are always either 1 or -1, for all fields and regions examined. These results appear less realistic than the OpenGGCM results obtained in Section~\ref{sect:OpenGGCM}, where DOI values have larger variability. The Tsyganenko models differ with respect to OpenGGCM in two fundamental aspects, in that they  (a) empirically reconstruct the magnetospheric magnetic field from an array of observations and (b) that they are not time-dependent. Either of these two aspects can contribute to the unrealistically high correlations we observe. Investigations on other models, and specifically on empirical, time-dependent models, will possibly help disentangle the role of these two aspects. At this stage of the investigation, we advance the hypothesis that time-dependent models may be better suited than time-independent models as background models for Data Assimilation techniques.

Third, with this analysis we have highlighted a possible path for future, targeted improvements of global heliospheric models used, among other things, for simulations of CME propagation in the heliosphere. It has long been known that one of the critical aspects of the simulation of CME arrival time is the estimation of the physical parameters to use as initial conditions in the simulations. While some parameters can be easily estimated from remote sensing, others are more difficult to determine properly and their variability affects the accuracy of the forecast (Falkenberg et al., 2010). In this paper, we have shown that DOI analysis could constitute an important stage of a model analysis effort aimed at clarifying which aspects of a model should be prioritized in order to obtain more accurate simulations of CME propagation.
 
In this study, as a first step, we show DOI maps obtained from the correlations of a single variable calculated between the variable at the observation point and the same variable in the domain under investigation. As demonstrated in Skandrani et al. (2014), cross-correlations can be used to find the influence of one variable upon another. 
 
The results of a DOI cross-correlation analysis can then be used to determine which quantities and areas in a simulation are most relevant in determining a certain observational quantity (such as the radial velocity of a CME in the case of CME propagation simulations). This analysis can then guide modelers on deciding which aspects of a model could be improved for more realistic results. It could help understanding, for example, if CME propagation in a model is mainly controlled by the background magnetic field configuration or by the properties of the CME itself at launch. In the first case, modeling efforts could be directed into accurate high resolution representation of the magnetic field configuration in the lower corona. In the second case, instead, modeling improvements could be focused on extracting better estimates of CME launch parameters (e.g. CME density, velocity, internal magnetic field configuration with respect to the background wind) from available observations.

The spatial correlations provided by DOI can also be of  particular interest in evaluating the effect of actual measurements done at positions different from the traditional L1, such as, for example, missions planned for L5 or missions closer to the Sun. 

Future work will extend this study to include temporal and cross correlations between different field components. This will further increase our knowledge of the models used to simulate such critical space weather processes.

The DOI analysis presented here can also be combined with an Observing System Simulation Experiment (OSSE), an approach already used in ionospheric and solar dynamo studies \citep{Hsu2018,Dikpati2017} to help provide a cost-effective approach to the evaluation of the potential impact of new observations.  OSSE requires that DA is already implemented and uses independently simulated ``data'' that are ingested into a different model or a different instance of the same model.  The effect of DA can then be investigated, albeit with caveats, since the ``data'' are not real. DOI analysis would obviate the need to have DA implemented, which can be very costly.  Instead, only ensemble runs with an unmodified model are required, and can provide a measure of the usefulness of a model and the available data for a specific situation.

\section*{Conflict of Interest Statement}
The authors declare that the research was conducted in the absence of any commercial or financial relationships that could be construed as a potential conflict of interest.

\section*{Author Contributions}
MEI, BL and DM created the ensembles using OpenGGCM, Tsyganenko and PLUTO respectively, as they appear in the main body of the article, processed the data and presented the results of the analysis and prepared the manuscript. JR provided OpenGGCM and contributed to the interpretation of the magnetosphere model results. JR, GL and SP provided support during the analysis and the preparation of the article.

\section*{Funding}
DM aknowledges support from AFRL (Air Force Research Laboratory)/USAF (US Air Force) project (AFRL Award No. FA9550-14-1-0375, 2014-2019) and partial support by UK STFC (Science and Technology Facilities Council) Consolidated Grant ST/S000240/1 (UCL-MSSL, University College London-Mullard Space Science Laboratory, Solar System). M.E.I.'s work is supported by an FWO (Fonds voor Wetenschappelijk Onderzoek – Vlaanderen) postdoctoral fellowship. 
BL, MEI, GL, JR acknowledge funding from the European Union’s Horizon 2020 research and innovation programme under grant agreement No 776262 (AIDA, Artificial Intelligence for Data Analysis, www.aida-space.eu). 
JR also acknowledges support through AFOSR grant FA9550-18-1-0483 and from the NASA/THEMIS mission through a subcontract from UC Berkeley.
SP and GL acknowledge funding from the European Union’s Horizon 2020 research and innovation programme under grant agreement No 870405. These results were also obtained in the framework of the projects C14/19/089  (C1 project Internal Funds KU Leuven), G.0A23.16N (FWO-Vlaanderen), C~90347 (ESA Prodex), and Belspo (Belgian Science Policy) BRAIN project BR/165/A2/CCSOM.

\section*{Acknowledgments}
The PLUTO simulations were performed using allocated time on the clusters Genius and Breniac. The computational resources and services used in this work were provided by the VSC (Flemish Supercomputer Center), funded by the Research Foundation Flanders (FWO) and the Flemish Government – department EWI. The OpenGGCM simulations were performed on the supercomputer Marconi-Broadwell (Cineca, Italy) under a PRACE allocation.
We acknowledge the NASA National Space Science Data Center, the Space Physics Data Facility, and the ACE Principal Investigator, Edward C. Stone of the California Institute of Technology, for usage of ACE data. We acknowledge use of NASA/GSFC's Space Physics Data Facility's OMNIWeb service, and OMNI data.

\section*{Supplemental Data}

\section*{Data Availability Statement}
The OMNIWeb database was used to get the solar wind conditions for the Tsyganenko runs. The ACE database was used to get the solar wind conditions for the OpenGGCM runs. The data used in the analysis are available upon reasonable request.

\bibliographystyle{frontiersinSCNS_ENG_HUMS} 

\bibliography{Paper_Frontiers_final}

\begin{thebibliography}{89}
\providecommand{\natexlab}[1]{#1}
\expandafter\ifx\csname urlstyle\endcsname\relax
  \providecommand{\doi}[1]{doi:\discretionary{}{}{}#1}\else
  \providecommand{\doi}{doi:\discretionary{}{}{}\begingroup
  \urlstyle{rm}\Url}\fi
\providecommand{\selectlanguage}[1]{\relax}
\providecommand{\bibAnnoteFile}[1]{%
  \IfFileExists{#1}{\begin{quotation}\noindent\textsc{Key:} #1\\
  \textsc{Annotation:}\ \input{#1}\end{quotation}}{}}
\providecommand{\bibAnnote}[2]{%
  \begin{quotation}\noindent\textsc{Key:} #1\\
  \textsc{Annotation:}\ #2\end{quotation}}

\bibitem[{Allen et~al.(1989)Allen, Sauer, Frank, and Reiff}]{allen1989effects}
Allen, J., Sauer, H., Frank, L., and Reiff, P. (1989).
\newblock Effects of the march 1989 solar activity.
\newblock \emph{Eos, Transactions American Geophysical Union} 70, 1479--1488
\bibAnnoteFile{allen1989effects}

\bibitem[{Altschuler and Newkirk(1969)}]{altschuler1969magnetic}
Altschuler, M.~D. and Newkirk, G. (1969).
\newblock Magnetic fields and the structure of the solar corona.
\newblock \emph{Solar Physics} 9, 131--149
\bibAnnoteFile{altschuler1969magnetic}

\bibitem[{Anderson et~al.(2017)Anderson, Korth, Welling, Merkin, Wiltberger,
  Raeder et~al.}]{Anderson2017CO}
Anderson, B.~J., Korth, H., Welling, D.~T., Merkin, V.~G., Wiltberger, M.~J.,
  Raeder, J., et~al. (2017).
\newblock Comparison of predictive estimates of high-latitude electrodynamics
  with observations of global-scale birkeland currents.
\newblock \emph{Space Weather} 15, 352--373.
\newblock \doi{10.1002/2016sw001529}
\bibAnnoteFile{Anderson2017CO}

\bibitem[{Anderson(2007)}]{anderson2007exploring}
Anderson, J.~L. (2007).
\newblock Exploring the need for localization in ensemble data assimilation
  using a hierarchical ensemble filter.
\newblock \emph{Physica D: Nonlinear Phenomena} 230, 99--111
\bibAnnoteFile{anderson2007exploring}

\bibitem[{Angelopoulos et~al.(1998)Angelopoulos, Carlson, Curtis, Harvey, Lin,
  Mozer et~al.}]{angelopoulos98a}
Angelopoulos, V., Carlson, C.~W., Curtis, D.~W., Harvey, P., Lin, R.~P., Mozer,
  F.~S., et~al. (1998).
\newblock On the necessity and feasability of a equatorial magnetospheric
  constellation.
\newblock In \emph{Science Closure and Enabling Technologies for Constellation
  Class Missions}, eds. V.~Angelopoulos and P.~V. Panetta (University of
  California, Berkeley, and NASA Goddard Space Flight Center). 14
\bibAnnoteFile{angelopoulos98a}

\bibitem[{Arge et~al.(2010)Arge, Henney, Koller, Compeau, Young, MacKenzie
  et~al.}]{arge2010air}
Arge, C.~N., Henney, C.~J., Koller, J., Compeau, C.~R., Young, S., MacKenzie,
  D., et~al. (2010).
\newblock Air force data assimilative photospheric flux transport (adapt)
  model.
\newblock In \emph{AIP Conference Proceedings} (AIP), vol. 1216, 343--346
\bibAnnoteFile{arge2010air}

\bibitem[{Bennett(1992)}]{bennett1992inverse}
Bennett, A.~F. (1992).
\newblock \emph{Inverse methods in physical oceanography} (Cambridge university
  press)
\bibAnnoteFile{bennett1992inverse}

\bibitem[{Berchem et~al.(1995)Berchem, Raeder, and Ashour-Abdalla}]{berchem95l}
Berchem, J., Raeder, J., and Ashour-Abdalla, M. (1995).
\newblock Reconnection at the magnetospheric boundary: Results from global
  {{MHD}} simulations.
\newblock In \emph{Physics of the Magnetopause}, eds. B.~U. Sonnerup and
  P.~Song. vol.~90 of \emph{AGU Geophysical Monograph}, 205
\bibAnnoteFile{berchem95l}

\bibitem[{Bishop and Hodyss(2007)}]{bishop2007flow}
Bishop, C.~H. and Hodyss, D. (2007).
\newblock Flow-adaptive moderation of spurious ensemble correlations and its
  use in ensemble-based data assimilation.
\newblock \emph{Quarterly Journal of the Royal Meteorological Society: A
  journal of the atmospheric sciences, applied meteorology and physical
  oceanography} 133, 2029--2044
\bibAnnoteFile{bishop2007flow}

\bibitem[{Bishop et~al.(2001)Bishop, Welch et~al.}]{bishop2001introduction}
Bishop, G., Welch, G., et~al. (2001).
\newblock An introduction to the kalman filter.
\newblock \emph{Proc of SIGGRAPH, Course} 8, 41
\bibAnnoteFile{bishop2001introduction}

\bibitem[{Bothmer and Daglis(2007)}]{bothmer2007space}
Bothmer, V. and Daglis, I.~A. (2007).
\newblock \emph{Space weather: physics and effects} (Springer Science \&
  Business Media)
\bibAnnoteFile{bothmer2007space}

\bibitem[{Bouttier and Courtier(1999)}]{bouttier1999data}
Bouttier, F. and Courtier, P. (1999).
\newblock Data assimilation concepts and methods. meteorological training
  course lecture series. ecmwf.
\newblock \emph{Reading, UK}
\bibAnnoteFile{bouttier1999data}

\bibitem[{Camporeale(2019)}]{camporeale2019challenge}
Camporeale, E. (2019).
\newblock The challenge of machine learning in space weather: Nowcasting and
  forecasting.
\newblock \emph{Space Weather} 17, 1166--1207
\bibAnnoteFile{camporeale2019challenge}

\bibitem[{{Chan{\'e}} et~al.(2005){Chan{\'e}}, {Jacobs}, {van der Holst},
  {Poedts}, and {Kimpe}}]{Chane2005}
{Chan{\'e}}, E., {Jacobs}, C., {van der Holst}, B., {Poedts}, S., and {Kimpe},
  D. (2005).
\newblock {On the effect of the initial magnetic polarity and of the background
  wind on the evolution of CME shocks}.
\newblock \emph{\aap} 432, 331--339.
\newblock \doi{10.1051/0004-6361:20042005}
\bibAnnoteFile{Chane2005}

\bibitem[{{Chan{\'e}} et~al.(2008){Chan{\'e}}, {Poedts}, and {van der
  Holst}}]{Chane2008}
{Chan{\'e}}, E., {Poedts}, S., and {van der Holst}, B. (2008).
\newblock {On the combination of ACE data with numerical simulations to
  determine the initial characteristics of a CME}.
\newblock \emph{\aap} 492, L29--L32.
\newblock \doi{10.1051/0004-6361:200811022}
\bibAnnoteFile{Chane2008}

\bibitem[{Connor et~al.(2016)Connor, Zesta, Fedrizzi, Shi, Raeder, Codrescu
  et~al.}]{Connor2016MO}
Connor, H.~K., Zesta, E., Fedrizzi, M., Shi, Y., Raeder, J., Codrescu, M.~V.,
  et~al. (2016).
\newblock Modeling the ionosphere-thermosphere response to a geomagnetic storm
  using physics-based magnetospheric energy input: {OpenGGCM}-{CTIM} results.
\newblock \emph{Journal of Space Weather and Space Climate} 6, A25.
\newblock \doi{10.1051/swsc/2016019}
\bibAnnoteFile{Connor2016MO}

\bibitem[{Cramer et~al.(2017)Cramer, Raeder, Toffoletto, Gilson, and
  Hu}]{cramer2017plasma}
Cramer, W.~D., Raeder, J., Toffoletto, F., Gilson, M., and Hu, B. (2017).
\newblock Plasma sheet injections into the inner magnetosphere: Two-way coupled
  openggcm-rcm model results.
\newblock \emph{Journal of Geophysical Research: Space Physics} 122, 5077--5091
\bibAnnoteFile{cramer2017plasma}

\bibitem[{{Dikpati}(2017)}]{Dikpati2017}
{Dikpati}, M. (2017).
\newblock {Ensemble Kalman Filter Data Assimilation in a Solar Dynamo Model}.
\newblock In \emph{AGU Fall Meeting Abstracts}. vol. 2017, SM14A--05
\bibAnnoteFile{Dikpati2017}

\bibitem[{Dorelli(2004)}]{Dorelli2004AN}
Dorelli, J.~C. (2004).
\newblock A new look at driven magnetic reconnection at the terrestrial
  subsolar magnetopause.
\newblock \emph{Journal of Geophysical Research} 109.
\newblock \doi{10.1029/2004ja010458}
\bibAnnoteFile{Dorelli2004AN}

\bibitem[{Eastwood et~al.(2017)Eastwood, Biffis, Hapgood, Green, Bisi, Bentley
  et~al.}]{Eastwood2017}
Eastwood, J.~P., Biffis, E., Hapgood, M.~A., Green, L., Bisi, M.~M., Bentley,
  R.~D., et~al. (2017).
\newblock The economic impact of space weather: Where do we stand?
\newblock \emph{Risk Analysis} 37, 206--218.
\newblock \doi{10.1111/risa.12765}
\bibAnnoteFile{Eastwood2017}

\bibitem[{Echevin et~al.(2000)Echevin, De~Mey, and
  Evensen}]{echevin2000horizontal}
Echevin, V., De~Mey, P., and Evensen, G. (2000).
\newblock Horizontal and vertical structure of the representer functions for
  sea surface measurements in a coastal circulation model.
\newblock \emph{Journal of physical oceanography} 30, 2627--2635
\bibAnnoteFile{echevin2000horizontal}

\bibitem[{Egbert and Bennett(1996)}]{egbert1996data}
[Dataset] Egbert, G. and Bennett, A. (1996).
\newblock Data assimilation methods for ocean tides, modern approaches to data
  assimilation in ocean modeling, edited by: Malonotte-rizzoli, p
\bibAnnoteFile{egbert1996data}

\bibitem[{Evensen(2009)}]{evensen2009data}
Evensen, G. (2009).
\newblock \emph{Data assimilation: the ensemble Kalman filter} (Springer
  Science \& Business Media)
\bibAnnoteFile{evensen2009data}

\bibitem[{Fortin et~al.(2014)Fortin, Abaza, Anctil, and
  Turcotte}]{fortin2014should}
Fortin, V., Abaza, M., Anctil, F., and Turcotte, R. (2014).
\newblock Why should ensemble spread match the rmse of the ensemble mean?
\newblock \emph{Journal of Hydrometeorology} 15, 1708--1713
\bibAnnoteFile{fortin2014should}

\bibitem[{Ge et~al.(2011)Ge, Raeder, Angelopoulos, Gilson, and Runov}]{Ge2011b}
Ge, Y.~S., Raeder, J., Angelopoulos, V., Gilson, M.~L., and Runov, A. (2011).
\newblock Interaction of dipolarization fronts within multiple bursty bulk
  flows in global {MHD} simulations of a substorm on 27 {February} 2009.
\newblock \emph{\jgr} 116, A00I23
\bibAnnoteFile{Ge2011b}

\bibitem[{Ghil and Malanotte-Rizzoli(1991)}]{ghil1991data}
Ghil, M. and Malanotte-Rizzoli, P. (1991).
\newblock Data assimilation in meteorology and oceanography.
\newblock In \emph{Advances in geophysics} (Elsevier), vol.~33. 141--266
\bibAnnoteFile{ghil1991data}

\bibitem[{{Hsu} et~al.(2018){Hsu}, {Matsuo}, and {Liu}}]{Hsu2018}
{Hsu}, C.~T., {Matsuo}, T., and {Liu}, J.~Y. (2018).
\newblock {Impact of Assimilating the FORMOSAT-3/COSMIC and FORMOSAT-7/COSMIC-2
  RO Data on the Midlatitude and Low-Latitude Ionospheric Specification}.
\newblock \emph{Earth and Space Science} 5, 875--890.
\newblock \doi{10.1029/2018EA000447}
\bibAnnoteFile{Hsu2018}

\bibitem[{Huang et~al.(2006)Huang, Spence, Lyon, Toffoletto, Singer, and
  Sazykin}]{huang2006storm}
Huang, C.-L., Spence, H.~E., Lyon, J.~G., Toffoletto, F.~R., Singer, H.~J., and
  Sazykin, S. (2006).
\newblock Storm-time configuration of the inner magnetosphere:
  Lyon-fedder-mobarry mhd code, tsyganenko model, and goes observations.
\newblock \emph{Journal of Geophysical Research: Space Physics} 111
\bibAnnoteFile{huang2006storm}

\bibitem[{Innocenti et~al.(2011)Innocenti, Lapenta, Vr{\v{s}}nak, Crespon,
  Skandrani, Temmer et~al.}]{innocenti2011improved}
Innocenti, M.~E., Lapenta, G., Vr{\v{s}}nak, B., Crespon, F., Skandrani, C.,
  Temmer, M., et~al. (2011).
\newblock Improved forecasts of solar wind parameters using the kalman filter.
\newblock \emph{Space Weather} 9, 1--15
\bibAnnoteFile{innocenti2011improved}

\bibitem[{Kalman(1960)}]{kalman1960new}
Kalman, R.~E. (1960).
\newblock A new approach to linear filtering and prediction problems.
\newblock \emph{Journal of basic Engineering} 82, 35--45
\bibAnnoteFile{kalman1960new}

\bibitem[{Kalnay(2003)}]{kalnay2003atmospheric}
Kalnay, E. (2003).
\newblock \emph{Atmospheric modeling, data assimilation and predictability}
  (Cambridge university press)
\bibAnnoteFile{kalnay2003atmospheric}

\bibitem[{{Keppens} and {Goedbloed}(2000)}]{Keppens2000}
{Keppens}, R. and {Goedbloed}, J.~P. (2000).
\newblock {Stellar Winds, Dead Zones, and Coronal Mass Ejections}.
\newblock \emph{\apj} 530, 1036--1048.
\newblock \doi{10.1086/308395}
\bibAnnoteFile{Keppens2000}

\bibitem[{King and Papitashvili(2005)}]{king2005solar}
King, J. and Papitashvili, N. (2005).
\newblock Solar wind spatial scales in and comparisons of hourly wind and ace
  plasma and magnetic field data.
\newblock \emph{Journal of Geophysical Research: Space Physics} 110
\bibAnnoteFile{king2005solar}

\bibitem[{Kondrashov et~al.(2007)Kondrashov, Shprits, Ghil, and
  Thorne}]{kondrashov2007kalman}
Kondrashov, D., Shprits, Y., Ghil, M., and Thorne, R. (2007).
\newblock A kalman filter technique to estimate relativistic electron lifetimes
  in the outer radiation belt.
\newblock \emph{Journal of Geophysical Research: Space Physics} 112
\bibAnnoteFile{kondrashov2007kalman}

\bibitem[{{Lang} et~al.(2020){Lang}, {Owens}, and {Lawless}}]{Lang2020}
{Lang}, M., {Owens}, M., and {Lawless}, A. (2020).
\newblock {Improving solar wind forecasts using data assimilation}.
\newblock In \emph{EGU General Assembly Conference Abstracts}. EGU General
  Assembly Conference Abstracts, 10909
\bibAnnoteFile{Lang2020}

\bibitem[{{Lang} and {Owens}(2019)}]{Lang2019}
{Lang}, M. and {Owens}, M.~J. (2019).
\newblock {A Variational Approach to Data Assimilation in the Solar Wind}.
\newblock \emph{Space Weather} 17, 59--83.
\newblock \doi{10.1029/2018SW001857}
\bibAnnoteFile{Lang2019}

\bibitem[{Lang and Owens(2019)}]{lang2019variational}
Lang, M. and Owens, M.~J. (2019).
\newblock A variational approach to data assimilation in the solar wind.
\newblock \emph{Space Weather} 17, 59--83
\bibAnnoteFile{lang2019variational}

\bibitem[{Laperre et~al.(2020)Laperre, Amaya, and Lapenta}]{laperre2020dynamic}
Laperre, B., Amaya, J., and Lapenta, G. (2020).
\newblock Dynamic time warping as a new evaluation for dst forecast with
  machine learning.
\newblock \emph{arXiv preprint arXiv:2006.04667}
\bibAnnoteFile{laperre2020dynamic}

\bibitem[{Lavraud et~al.(2016)Lavraud, Liu, Segura, He, Qin, Temmer
  et~al.}]{Lavraud2016}
Lavraud, B., Liu, Y., Segura, K., He, J., Qin, G., Temmer, M., et~al. (2016).
\newblock A small mission concept to the sun–earth lagrangian l5 point for
  innovative solar, heliospheric and space weather science.
\newblock \emph{Journal of Atmospheric and Solar-Terrestrial Physics} 146, 171
  -- 185.
\newblock \doi{https://doi.org/10.1016/j.jastp.2016.06.004}
\bibAnnoteFile{Lavraud2016}

\bibitem[{Le~Dimet and Talagrand(1986)}]{le1986variational}
Le~Dimet, F.-X. and Talagrand, O. (1986).
\newblock Variational algorithms for analysis and assimilation of
  meteorological observations: theoretical aspects.
\newblock \emph{Tellus A: Dynamic Meteorology and Oceanography} 38, 97--110
\bibAnnoteFile{le1986variational}

\bibitem[{Luhmann et~al.(2004)Luhmann, Solomon, Linker, Lyon, Mikic, Odstrcil
  et~al.}]{luhmann2004coupled}
Luhmann, J.~G., Solomon, S.~C., Linker, J.~A., Lyon, J.~G., Mikic, Z.,
  Odstrcil, D., et~al. (2004).
\newblock Coupled model simulation of a sun-to-earth space weather event.
\newblock \emph{Journal of atmospheric and solar-terrestrial physics} 66,
  1243--1256
\bibAnnoteFile{luhmann2004coupled}

\bibitem[{McComas et~al.(1998)McComas, Bame, Barraclough, Feldman, Funsten,
  Gosling et~al.}]{McComas1998}
McComas, D.~J., Bame, S.~J., Barraclough, B.~L., Feldman, W.~C., Funsten,
  H.~O., Gosling, J.~T., et~al. (1998).
\newblock Ulysses' return to the slow solar wind.
\newblock \emph{Geophysical Research Letters} 25, 1--4.
\newblock \doi{10.1029/97GL03444}
\bibAnnoteFile{McComas1998}

\bibitem[{Mendoza et~al.(2006)Mendoza, De~Moor, and
  Bernstein}]{mendoza2006data}
Mendoza, O.~B., De~Moor, B., and Bernstein, D. (2006).
\newblock Data assimilation for magnetohydrodynamics systems.
\newblock \emph{Journal of Computational and Applied Mathematics} 189, 242--259
\bibAnnoteFile{mendoza2006data}

\bibitem[{{Mignone} et~al.(2007){Mignone}, {Bodo}, {Massaglia}, {Matsakos},
  {Tesileanu}, {Zanni} et~al.}]{Mignone2007}
{Mignone}, A., {Bodo}, G., {Massaglia}, S., {Matsakos}, T., {Tesileanu}, O.,
  {Zanni}, C., et~al. (2007).
\newblock {PLUTO: A Numerical Code for Computational Astrophysics}.
\newblock \emph{\apjs} 170, 228--242.
\newblock \doi{10.1086/513316}
\bibAnnoteFile{Mignone2007}

\bibitem[{{Mignone} et~al.(2012){Mignone}, {Zanni}, {Tzeferacos}, {van
  Straalen}, {Colella}, and {Bodo}}]{Mignone2012}
{Mignone}, A., {Zanni}, C., {Tzeferacos}, P., {van Straalen}, B., {Colella},
  P., and {Bodo}, G. (2012).
\newblock {The PLUTO Code for Adaptive Mesh Computations in Astrophysical Fluid
  Dynamics}.
\newblock \emph{\apjs} 198, 7.
\newblock \doi{10.1088/0067-0049/198/1/7}
\bibAnnoteFile{Mignone2012}

\bibitem[{Moretto et~al.(2006)Moretto, Vennerstrom, Olsen, Rastaetter, and
  Raeder}]{moretto2005}
Moretto, T., Vennerstrom, S., Olsen, N., Rastaetter, L., and Raeder, J. (2006).
\newblock Using global magnetospheric models for simulation and interpretation
  of {SWARM} external field measurements.
\newblock \emph{Earth, Planets, Space} 58, 439--449
\bibAnnoteFile{moretto2005}

\bibitem[{Nikolić(2019)}]{nikolic2019}
Nikolić, L. (2019).
\newblock On solutions of the pfss model with gong synoptic maps for
  2006–2018.
\newblock \emph{Space Weather} 17, 1293--1311.
\newblock \doi{10.1029/2019SW002205}
\bibAnnoteFile{nikolic2019}

\bibitem[{{Owens} et~al.(2020){Owens}, {Lang}, {Barnard}, {Riley}, {Ben-Nun},
  {Scott} et~al.}]{Owens2020}
{Owens}, M., {Lang}, M., {Barnard}, L., {Riley}, P., {Ben-Nun}, M., {Scott},
  C.~J., et~al. (2020).
\newblock {A Computationally Efficient, Time-Dependent Model of the Solar Wind
  for Use as a Surrogate to Three-Dimensional Numerical Magnetohydrodynamic
  Simulations}.
\newblock \emph{\solphys} 295, 43.
\newblock \doi{10.1007/s11207-020-01605-3}
\bibAnnoteFile{Owens2020}

\bibitem[{Pevtsov et~al.(2020)Pevtsov, Petrie, MacNeice, and
  Virtanen}]{Pevtsov2020}
Pevtsov, A.~A., Petrie, G., MacNeice, P., and Virtanen, I.~I. (2020).
\newblock Effect of additional magnetograph observations from different
  lagrangian points in sun-earth system on predicted properties of quasi-steady
  solar wind at 1 au.
\newblock \emph{Space Weather} 18, e2020SW002448.
\newblock \doi{10.1029/2020SW002448}.
\newblock E2020SW002448 10.1029/2020SW002448
\bibAnnoteFile{Pevtsov2020}

\bibitem[{Plunkett(2005)}]{plunkett2005extreme}
Plunkett, S. (2005).
\newblock \emph{The extreme solar storms of October to November 2003}.
\newblock Tech. rep., NAVAL RESEARCH LAB WASHINGTON DC SPACE SCIENCE DIV
\bibAnnoteFile{plunkett2005extreme}

\bibitem[{Pulkkinen et~al.(2005)Pulkkinen, Lindahl, Viljanen, and
  Pirjola}]{pulkkinen2005geomagnetic}
Pulkkinen, A., Lindahl, S., Viljanen, A., and Pirjola, R. (2005).
\newblock Geomagnetic storm of 29--31 october 2003: Geomagnetically induced
  currents and their relation to problems in the swedish high-voltage power
  transmission system.
\newblock \emph{Space Weather} 3
\bibAnnoteFile{pulkkinen2005geomagnetic}

\bibitem[{Raeder(2003)}]{raeder03t}
Raeder, J. (2003).
\newblock {Global} {Magnetohydrodynamics} -- {A} {Tutorial}.
\newblock In \emph{{Space} {Plasma} {Simulation}}, eds. {J.~B\"uchner}, C.~T.
  Dum, and M.~Scholer (Berlin Heidelberg New York: Springer Verlag).
\newblock \doi{10.1007/3-540-36530-3_11}
\bibAnnoteFile{raeder03t}

\bibitem[{Raeder(2006)}]{Raeder2006FL}
Raeder, J. (2006).
\newblock Flux transfer events: 1. generation mechanism for strong southward
  {IMF}.
\newblock \emph{Annales Geophysicae} 24, 381--392.
\newblock \doi{10.5194/angeo-24-381-2006}
\bibAnnoteFile{Raeder2006FL}

\bibitem[{Raeder and Angelopoulos(1998)}]{raeder98e}
Raeder, J. and Angelopoulos, V. (1998).
\newblock Using global simulations of the magnetosphere for multi-satellite
  mission planning and analysis,.
\newblock In \emph{Science Closure and Enabling Technologies for Constellation
  Class Missions}, eds. V.~Angelopoulos and P.~V. Panetta (University of
  California, Berkeley, and NASA Goddard Space Flight Center). 78
\bibAnnoteFile{raeder98e}

\bibitem[{{Raeder} et~al.(2017){Raeder}, {Cramer}, {Germaschewski}, and
  {Jensen}}]{Raeder2017}
{Raeder}, J., {Cramer}, W.~D., {Germaschewski}, K., and {Jensen}, J. (2017).
\newblock {Using OpenGGCM to Compute and Separate Magnetosphere Magnetic
  Perturbations Measured on Board Low Earth Orbiting Satellites}.
\newblock \emph{\ssr} 206, 601--620.
\newblock \doi{10.1007/s11214-016-0304-x}
\bibAnnoteFile{Raeder2017}

\bibitem[{Raeder and Lu(2005)}]{Raeder2005PO}
Raeder, J. and Lu, G. (2005).
\newblock Polar cap potential saturation during large geomagnetic storms.
\newblock \emph{Advances in Space Research} 36, 1804--1808.
\newblock \doi{10.1016/j.asr.2004.05.010}
\bibAnnoteFile{Raeder2005PO}

\bibitem[{Raeder et~al.(2001{\natexlab{a}})Raeder, McPherron, Frank, Paterson,
  Sigwarth, Lu et~al.}]{raeder00c}
Raeder, J., McPherron, R.~L., Frank, L.~A., Paterson, W.~R., Sigwarth, J.~B.,
  Lu, G., et~al. (2001{\natexlab{a}}).
\newblock Global simulation of the geospace environment modeling substorm
  challenge event.
\newblock \emph{\jgr} 106, 381
\bibAnnoteFile{raeder00c}

\bibitem[{Raeder et~al.(2001{\natexlab{b}})Raeder, Wang, Fuller-Rowell, and
  Singer}]{raeder01a}
Raeder, J., Wang, Y.~L., Fuller-Rowell, T.~J., and Singer, H.~J.
  (2001{\natexlab{b}}).
\newblock Global simulation of space weather effects of the {Bastille} {Day}
  storm.
\newblock \emph{Sol. Phys.} 204, 325
\bibAnnoteFile{raeder01a}

\bibitem[{Raeder et~al.(2010)Raeder, Zhu, Ge, and Siscoe}]{raeder2010a}
Raeder, J., Zhu, P., Ge, Y., and Siscoe, G.~L. (2010).
\newblock {OpenGGCM} simulation of a substorm: {Axial} tail instability and
  ballooning mode preceding substorm onset.
\newblock \emph{\jgr} 115, A00l16
\bibAnnoteFile{raeder2010a}

\bibitem[{Richardson and Cane(2010)}]{richardson2010near}
Richardson, I.~G. and Cane, H.~V. (2010).
\newblock Near-earth interplanetary coronal mass ejections during solar cycle
  23 (1996--2009): Catalog and summary of properties.
\newblock \emph{Solar Physics} 264, 189--237
\bibAnnoteFile{richardson2010near}

\bibitem[{Rigler et~al.(2004)Rigler, Baker, Weigel, Vassiliadis, and
  Klimas}]{rigler2004adaptive}
Rigler, E., Baker, D., Weigel, R., Vassiliadis, D., and Klimas, A. (2004).
\newblock Adaptive linear prediction of radiation belt electrons using the
  kalman filter.
\newblock \emph{Space Weather} 2, 1--9
\bibAnnoteFile{rigler2004adaptive}

\bibitem[{Sakov and Bertino(2011)}]{sakov2011relation}
Sakov, P. and Bertino, L. (2011).
\newblock Relation between two common localisation methods for the enkf.
\newblock \emph{Computational Geosciences} 15, 225--237
\bibAnnoteFile{sakov2011relation}

\bibitem[{Schatten(1971)}]{schatten1971current}
Schatten, K.~H. (1971).
\newblock Current sheet magnetic model for the solar corona
\bibAnnoteFile{schatten1971current}

\bibitem[{Schatten et~al.(1969)Schatten, Wilcox, and Ness}]{schatten1969model}
Schatten, K.~H., Wilcox, J.~M., and Ness, N.~F. (1969).
\newblock A model of interplanetary and coronal magnetic fields.
\newblock \emph{Solar Physics} 6, 442--455
\bibAnnoteFile{schatten1969model}

\bibitem[{Schrijver and DeRosa(2003)}]{schrijver2003photospheric}
Schrijver, C.~J. and DeRosa, M.~L. (2003).
\newblock Photospheric and heliospheric magnetic fields.
\newblock \emph{Solar Physics} 212, 165--200
\bibAnnoteFile{schrijver2003photospheric}

\bibitem[{Schunk et~al.(2004)Schunk, Scherliess, Sojka, Thompson, Anderson,
  Codrescu et~al.}]{schunk2004global}
Schunk, R.~W., Scherliess, L., Sojka, J.~J., Thompson, D.~C., Anderson, D.~N.,
  Codrescu, M., et~al. (2004).
\newblock Global assimilation of ionospheric measurements (gaim).
\newblock \emph{Radio Science} 39
\bibAnnoteFile{schunk2004global}

\bibitem[{Shi et~al.(2014)Shi, Hartinger, Angelopoulos, Tian, Fu, Zong
  et~al.}]{Shi2014SO}
Shi, Q.~Q., Hartinger, M., Angelopoulos, V., Tian, A., Fu, S., Zong, Q.-G.,
  et~al. (2014).
\newblock Solar wind pressure pulse-driven magnetospheric vortices and their
  global consequences.
\newblock \emph{Journal of Geophysical Research: Space Physics} 119,
  4274--4280.
\newblock \doi{10.1002/2013ja019551}
\bibAnnoteFile{Shi2014SO}

\bibitem[{Siscoe et~al.(2004)Siscoe, Baker, Weigel, Hughes, and
  Spence}]{siscoe2004roles}
Siscoe, G., Baker, D., Weigel, R., Hughes, J., and Spence, H. (2004).
\newblock Roles of empirical modeling within cism.
\newblock \emph{Journal of atmospheric and solar-terrestrial physics} 66,
  1481--1489
\bibAnnoteFile{siscoe2004roles}

\bibitem[{Skandrani et~al.(2014)Skandrani, Innocenti, Bettarini, Crespon,
  Lamouroux, and Lapenta}]{skandrani2014flip}
Skandrani, C., Innocenti, M.~E., Bettarini, L., Crespon, F., Lamouroux, J., and
  Lapenta, G. (2014).
\newblock Flip-mhd-based model sensitivity analysis.
\newblock \emph{Nonlinear processes in geophysics} 21, 539--553
\bibAnnoteFile{skandrani2014flip}

\bibitem[{Stone et~al.(1998)Stone, Frandsen, Mewaldt, Christian, Margolies,
  Ormes et~al.}]{stone1998advanced}
Stone, E.~C., Frandsen, A., Mewaldt, R., Christian, E., Margolies, D., Ormes,
  J., et~al. (1998).
\newblock The advanced composition explorer.
\newblock \emph{Space Science Reviews} 86, 1--22
\bibAnnoteFile{stone1998advanced}

\bibitem[{Thomsen et~al.(1996)Thomsen, McComas, Reeves, and
  Weiss}]{thomsen1996observational}
Thomsen, M., McComas, D., Reeves, G., and Weiss, L. (1996).
\newblock An observational test of the tsyganenko (t89a) model of the
  magnetospheric field.
\newblock \emph{Journal of Geophysical Research: Space Physics} 101,
  24827--24836
\bibAnnoteFile{thomsen1996observational}

\bibitem[{Toffoletto et~al.(2001)Toffoletto, Spiro, Wolf, Birn, and
  Hesse}]{toffoletto2001modeling}
Toffoletto, F., Spiro, R., Wolf, R., Birn, J., and Hesse, M. (2001).
\newblock Modeling inner magnetospheric electrodynamics.
\newblock \emph{Washington DC American Geophysical Union Geophysical Monograph
  Series} 125, 265--272
\bibAnnoteFile{toffoletto2001modeling}

\bibitem[{T{\'o}th et~al.(2005)T{\'o}th, Sokolov, Gombosi, Chesney, Clauer,
  De~Zeeuw et~al.}]{toth2005space}
T{\'o}th, G., Sokolov, I.~V., Gombosi, T.~I., Chesney, D.~R., Clauer, C.~R.,
  De~Zeeuw, D.~L., et~al. (2005).
\newblock Space weather modeling framework: A new tool for the space science
  community.
\newblock \emph{Journal of Geophysical Research: Space Physics} 110
\bibAnnoteFile{toth2005space}

\bibitem[{Tsurutani et~al.(2006)Tsurutani, Gonzalez, Gonzalez, Guarnieri,
  Gopalswamy, Grande et~al.}]{tsurutani2006corotating}
Tsurutani, B.~T., Gonzalez, W.~D., Gonzalez, A.~L., Guarnieri, F.~L.,
  Gopalswamy, N., Grande, M., et~al. (2006).
\newblock Corotating solar wind streams and recurrent geomagnetic activity: A
  review.
\newblock \emph{Journal of Geophysical Research: Space Physics} 111
\bibAnnoteFile{tsurutani2006corotating}

\bibitem[{Tsyganenko(1987)}]{tsyganenko1987global}
Tsyganenko, N. (1987).
\newblock Global quantitative models of the geomagnetic field in the cislunar
  magnetosphere for different disturbance levels.
\newblock \emph{Planetary and space science} 35, 1347--1358
\bibAnnoteFile{tsyganenko1987global}

\bibitem[{Tsyganenko(1989)}]{tsyganenko1989magnetospheric}
Tsyganenko, N. (1989).
\newblock A magnetospheric magnetic field model with a warped tail current
  sheet.
\newblock \emph{Planetary and Space Science} 37, 5--20
\bibAnnoteFile{tsyganenko1989magnetospheric}

\bibitem[{Tsyganenko(2002{\natexlab{a}})}]{tsyganenko2002modela}
Tsyganenko, N. (2002{\natexlab{a}}).
\newblock A model of the near magnetosphere with a dawn-dusk asymmetry 1.
  mathematical structure.
\newblock \emph{Journal of Geophysical Research: Space Physics} 107, SMP--12
\bibAnnoteFile{tsyganenko2002modela}

\bibitem[{Tsyganenko(2002{\natexlab{b}})}]{tsyganenko2002modelb}
Tsyganenko, N. (2002{\natexlab{b}}).
\newblock A model of the near magnetosphere with a dawn-dusk asymmetry 2.
  parameterization and fitting to observations.
\newblock \emph{Journal of Geophysical Research: Space Physics} 107, SMP--10
\bibAnnoteFile{tsyganenko2002modelb}

\bibitem[{Tsyganenko and Andreeva(2015)}]{tsyganenko2015forecasting}
Tsyganenko, N. and Andreeva, V. (2015).
\newblock A forecasting model of the magnetosphere driven by an optimal solar
  wind coupling function.
\newblock \emph{Journal of Geophysical Research: Space Physics} 120, 8401--8425
\bibAnnoteFile{tsyganenko2015forecasting}

\bibitem[{Tsyganenko et~al.(2003)Tsyganenko, Singer, and
  Kasper}]{tsyganenko2003storm}
Tsyganenko, N., Singer, H., and Kasper, J. (2003).
\newblock Storm-time distortion of the inner magnetosphere: How severe can it
  get?
\newblock \emph{Journal of Geophysical Research: Space Physics} 108
\bibAnnoteFile{tsyganenko2003storm}

\bibitem[{Tsyganenko and Sitnov(2005)}]{tsyganenko2005modeling}
Tsyganenko, N. and Sitnov, M. (2005).
\newblock Modeling the dynamics of the inner magnetosphere during strong
  geomagnetic storms.
\newblock \emph{Journal of Geophysical Research: Space Physics} 110
\bibAnnoteFile{tsyganenko2005modeling}

\bibitem[{Tsyganenko(1995)}]{tsyganenko1995modeling}
Tsyganenko, N.~A. (1995).
\newblock Modeling the earth's magnetospheric magnetic field confined within a
  realistic magnetopause.
\newblock \emph{Journal of Geophysical Research: Space Physics} 100, 5599--5612
\bibAnnoteFile{tsyganenko1995modeling}

\bibitem[{Tsyganenko(1996)}]{tsyganenko1996effects}
Tsyganenko, N.~A. (1996).
\newblock Effects of the solar wind conditions in the global magnetospheric
  configurations as deduced from data-based field models.
\newblock In \emph{International conference on substorms}. vol. 389, 181
\bibAnnoteFile{tsyganenko1996effects}

\bibitem[{Vennerstrom et~al.(2005)Vennerstrom, Moretto, Rastaetter, and
  Raeder}]{vennerstroe2005}
Vennerstrom, S., Moretto, T., Rastaetter, L., and Raeder, J. (2005).
\newblock Field-aligned currents during northward interplanetary field:
  {Morphology and causes}.
\newblock \emph{\jgr} 110, A06205, ~{\sf doi: 10.1029/2004JA010802}
\bibAnnoteFile{vennerstroe2005}

\bibitem[{Vourlidas(2015)}]{Vourlidas2015}
Vourlidas, A. (2015).
\newblock Mission to the sun-earth l5 lagrangian point: An optimal platform for
  space weather research.
\newblock \emph{Space Weather} 13, 197--201.
\newblock \doi{10.1002/2015SW001173}
\bibAnnoteFile{Vourlidas2015}

\bibitem[{Wang and Sheeley~Jr(1992)}]{wang1992potential}
Wang, Y.-M. and Sheeley~Jr, N. (1992).
\newblock On potential field models of the solar corona.
\newblock \emph{The Astrophysical Journal} 392, 310--319
\bibAnnoteFile{wang1992potential}

\bibitem[{Woodfield et~al.(2007)Woodfield, Dunlop, Holme, Davies, and
  Hapgood}]{woodfield2007comparison}
Woodfield, E., Dunlop, M., Holme, R., Davies, J., and Hapgood, M. (2007).
\newblock A comparison of cluster magnetic data with the tsyganenko 2001 model.
\newblock \emph{Journal of Geophysical Research: Space Physics} 112
\bibAnnoteFile{woodfield2007comparison}

\bibitem[{Zhou et~al.(2012)Zhou, Ge, Angelopoulos, Runov, Liang, Xing
  et~al.}]{Zhou2012DI}
Zhou, X.-Z., Ge, Y.~S., Angelopoulos, V., Runov, A., Liang, J., Xing, X.,
  et~al. (2012).
\newblock Dipolarization fronts and associated auroral activities: 2.
  acceleration of ions and their subsequent behavior.
\newblock \emph{Journal of Geophysical Research: Space Physics} 117, 1.
\newblock \doi{10.1029/2012ja017677}
\bibAnnoteFile{Zhou2012DI}

\bibitem[{Zhu et~al.(2009)Zhu, Raeder, Germaschewski, and Hegna}]{Zhu2009IN}
Zhu, P., Raeder, J., Germaschewski, K., and Hegna, C.~C. (2009).
\newblock Initiation of ballooning instability in the near-{Earth} plasma sheet
  prior to the 23 march 2007 {{THEMIS}} substorm expansion onset.
\newblock \emph{Annales Geophysicae} 27, 1129--1138.
\newblock \doi{10.5194/angeo-27-1129-2009}
\bibAnnoteFile{Zhu2009IN}

\end{thebibliography}

\end{document}


\onecolumn
\firstpage{1}

\title[Supplementary Material]{{\helveticaitalic{Supplementary Material}}}

\maketitle

\section{Supplementary Data}

We provide as supplementary material the following animations of simulations with OpenGGCM:

\begin{enumerate}
    \item Dynamic representation of the open magnetosphere scenario using OpenGGCM (shown in ReferenceSimVx.avi, panel \ref{fig: mov1}), showing the temporal variation of the $b_z$ component and reconnection events.
    \item Global evolution of the ensemble mean (MeanVx.avi, panel \ref{fig: mov2}), where the modified magnetopause-magnetotail reconnection pattern is visible.
    \item Evolution of the members of the ensemble using different values for the $v_x$ component. Two different values are tested, $|v_x| \sim 363$ km/s (LowerVx.avi, panel \ref{fig: mov3}) and $|v_x| \sim 583$ km/s (HigherVx.avi, panel \ref{fig: mov4}), resulting in two different magnetosheath patterns.
    \item The DOI movies for the Bx and Vx with observation points in the plasma sheet (panels \ref{fig: mov5} and \ref{fig: mov6}).
    \item The DOI movies for the Bx and Vx with observation points  in the magnetosheath (panels \ref{fig: mov7} and \ref{fig: mov8}).
\end{enumerate}

\begin{figure}
\centering
\includemedia[width=0.7\linewidth,height=0.45\linewidth,activate=onclick,
transparent,addresource=ReferenceSimVx.avi,flashvars={source=ReferenceSimVx.avi}
]{}{StrobeMediaPlayback.swf}
\caption{Temporal variation of $b_z$, reference run}\label{fig: mov1}
\end{figure}

\begin{figure}
\centering
\includemedia[width=0.7\linewidth,height=0.45\linewidth,activate=onclick,
transparent,addresource=MeanVx.avi,flashvars={source=MeanVx.avi}
]{}{StrobeMediaPlayback.swf}
\caption{Evolution of the ensemble mean}\label{fig: mov2}
\end{figure}

\begin{figure}
\centering
\includemedia[width=0.7\linewidth,height=0.45\linewidth,activate=onclick,
transparent,addresource=LowerVx.avi,flashvars={source=LowerVx.avi}
]{}{StrobeMediaPlayback.swf}
\caption{Evolution when $|v_x| \sim363$km/s}\label{fig: mov3}
\end{figure}

\begin{figure}
\centering
\includemedia[width=0.7\linewidth,height=0.45\linewidth,activate=onclick,
transparent,addresource=HigherVx.avi,flashvars={source=HigherVx.avi}
]{}{StrobeMediaPlayback.swf}
\caption{Evolution for  $|v_x| \sim583$km/s}\label{fig: mov4}
\end{figure}

\begin{figure}
\centering
\includemedia[width=0.7\linewidth,height=0.45\linewidth,activate=onclick,
transparent,addresource=DOI_bx_bx_pSheet.avi,flashvars={source=DOI_bx_bx_pSheet.avi}
]{}{StrobeMediaPlayback.swf}
\caption{DOI for Bx, observation point in the plasmasheet}\label{fig: mov5}
\end{figure}

\begin{figure}
\centering
\includemedia[width=0.7\linewidth,height=0.45\linewidth,activate=onclick,
transparent,addresource=DOI_vx_vx_pSheet.avi,flashvars={source=DOI_vx_vx_pSheet.avi}
]{}{StrobeMediaPlayback.swf}
\caption{DOI for Vx, observation point in the plasmasheet}\label{fig: mov6}
\end{figure}

\begin{figure}
\centering
\includemedia[width=0.7\linewidth,height=0.45\linewidth,activate=onclick,
transparent,addresource=DOI_bx_bx_MSheath.avi,flashvars={source=DOI_bx_bx_MSheath.avi}
]{}{StrobeMediaPlayback.swf}
\caption{DOI for Bx, observation point in the magnethosheath}\label{fig: mov7}
\end{figure}

\begin{figure}
\centering
\includemedia[width=0.7\linewidth,height=0.45\linewidth,activate=onclick,
transparent,addresource=DOI_vx_vx_MSheath.avi,flashvars={source=DOI_vx_vx_MSheath.avi}
]{}{StrobeMediaPlayback.swf}
\caption{DOI for Vx, observation point in the magnethosheath}\label{fig: mov8}
\end{figure}
